\documentclass[twocolumn,showpacs,preprintnumbers,amsmath,amssymb,pre]{revtex4}
\usepackage{graphicx,amsmath,amssymb,wrapfig,mathrsfs}

\newcommand{\beq}{\begin{eqnarray}}
\newcommand{\eeq}{\end{eqnarray}}
\newcommand{\eps}{\varepsilon}
\newcommand{\p}{\partial}

\begin{document}

\title{Dynamics of Free Surface Perturbations\\
Along an Annular Viscous Film}
\author{Linda B. Smolka}
\email{lsmolka@bucknell.edu}
\affiliation{\it Department of Mathematics, Bucknell University, Lewisburg,
PA 17837, USA}
\author{Justin North}
\affiliation{\it Department of Physics, Ohio State University, Columbus,
OH 43210-1117, USA}
\author{Bree K. Guerra}
\affiliation{\it Department of Physics, University of Texas at Austin, 
Austin, TX 78712-0264, USA}
\date{\today}

\begin{abstract}  
It is known that the free surface of an 
axisymmetric viscous film flowing down the outside of a thin vertical 
fiber under the influence of gravity becomes unstable to interfacial 
perturbations.  We present an experimental study using fluids with 
different densities, surface tensions and viscosities to investigate the 
growth and dynamics of these interfacial perturbations and to test the 
assumptions made by previous authors.  We find the initial perturbation 
growth is exponential followed by a slower phase as the amplitude and 
wavelength saturate in size.  Measurements of the perturbation growth for 
experiments conducted at low and moderate Reynolds numbers are 
compared to theoretical predictions developed from linear stability 
theory.  Excellent agreement is found between predictions from a 
long-wave Stokes flow model (Craster \& Matar, J. Fluid Mech. 
{\bf 553}, 85 (2006)) and data, while fair agreement is found 
between predictions from a moderate Reynolds number model 
(Sisoev {\it et al.}, Chem.~Eng.~Sci. {\bf 61}, 7279 (2006)) and 
data.  Furthermore, we find that a known transition in the longer-time
perturbation dynamics from unsteady to steady behavior at a critical 
flow rate, $Q_c,$ is correlated to a transition in the rate at which 
perturbations naturally form along the fiber.  For $Q<Q_c$ (steady case), 
the rate of perturbation formation is constant. As a result the position 
along the fiber where perturbations form is nearly fixed, and the spacing 
between consecutive perturbations remains constant as they travel 2~m 
down the fiber.  For $Q>Q_c$ (unsteady case), the rate of perturbation 
formation is modulated.  As a result the position along the fiber 
where perturbations form oscillates irregularly, and the initial speed 
and spacing between perturbations varies resulting in the coalescence 
of neighboring perturbations further down the fiber. 
\end{abstract}


\maketitle

\section{Introduction}
\label{sec-intro}

Coatings are commonly applied to the exterior of thin cylindrical
wires or fibers to provide protection and/or enhance performance
(e.g. electrical wire and fiber-optic cable).  Methods of coating 
include extruding a fiber through a die (die coating) or drawing 
a fiber from a liquid bath (dip coating) 
\cite{quere99,ruck02,deryck96,deryck98,shen02}.  During the coating 
process, a uniform liquid film can become unstable to interfacial 
perturbations that may develop further into droplets 
\cite{goren62,quere90}.  This effect, which detracts from 
the quality of a coating, has inspired a wide array of studies
on the formation and motion of perturbations on cylindrical fibers 
\cite{chang99,frenk92,goren62,kall94,lin75,lister06,quere90,zucch05}.

\begin{figure}[h]	
\begin{center}	

\includegraphics [width=3.4 in]{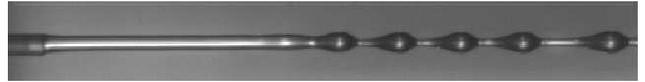}
													 
\caption{An annular film of viscous fluid flowing down the outside 
of a thin nylon fiber; the surrounding fluid is air.  Note that
the figure has been rotated by 90$^\text{o}$ with gravity acting towards
the right.  Perturbations develop along the free surface some 
distance down the fiber; once formed, these perturbations continue 
to travel down the fiber. Image length = 9.7~cm.}  

\label{f-anjet}										
\end{center}
							
\end{figure}

Fibers can also be coated by a continuously-fed axisymmetric fluid 
flow down the length of a vertical fiber (see Figure~\ref{f-anjet})
as has been examined in several analytical and experimental 
studies \cite{cras06,duprat07,kliak01,sisoev06,solo87,trif92}.  
The geometry of the unperturbed flow is an annular film with a fixed 
internal boundary and a free surface at the outer fluid-air interface.  
It is well known that the free surface of this annular film becomes 
unstable to interfacial perturbations, as shown in Figure~\ref{f-anjet}.  
Herein we present an experimental study on an annular viscous film with 
a particular focus on the initial formation and longer-time dynamics of 
perturbations along the film free surface.


A related problem to annular films is on the motion of cylindrical jets, 
which in contrast have no fixed internal boundary.  Analytical studies
on the motion and stability of inviscid and viscous jets dates back to 
the work of Plateau \cite{plat73}, Lord Rayleigh \cite{ray1879,ray1892}, 
Weber \cite{weber31} and Chandrasekhar \cite{chan61}.  It is 
known that capillary effects drive perturbation growth along the jet 
free surface, often referred to as the Plateau-Rayleigh instability.  
Analytical results developed from temporal linear stability theory 
\cite{chan61} were tested in experiments by Donnelly \& Glaberson 
\cite{donn66}, who found strong (fair) quantitative agreement between 
the theoretical and measured dispersion relation for an inviscid 
(viscous) jet.  The dynamics of free surface perturbations along 
cylindrical jets and annular films are, however, quite different.  
In the cylindrical case jet breakup occurs when the perturbations 
become sufficiently large \cite{donn66}, while in the annular case 
the large amplitude perturbations remain connected by a liquid film 
\cite{cras06,duprat07,kliak01}.

Recently, several theoretical studies have analyzed the temporal 
linear stability of an annular viscous film flowing down a vertical 
fiber in the Stokes \cite{cras06,kliak01} and moderate Reynolds number 
\cite{trif92,sisoev06} limits; the base flow in these studies is 
assumed to be a steady, unidirectional parallel flow.  Here we test 
these results by determining whether: (i) the base flow used in 
\cite{cras06,kliak01,trif92,sisoev06} matches the experimental flow; 
and (ii) the dispersion relations derived in the Stokes and moderate 
Reynolds number limits \cite{cras06,sisoev06} correctly predict 
the nascent growth of perturbations measured in low and moderate 
Reynolds number flows.

As free surface perturbations travel down a vertical fiber, many 
interesting phenomena occur \cite{duprat07,cras06,kliak01,sisoev06}.  
In experiments, Kliakhandler, Davis and Bankoff (KDB) observed three 
types of behavior far down the length of the fiber ($\approx$~2~m) 
\cite{kliak01}.  At the highest flow rate (regime a), the film between 
perturbations is thick and uniform, and faster moving perturbations 
collide into slower moving perturbations (unsteady behavior).  At an 
intermediate flow rate (regime b), the spacing, size and speed of the 
perturbations is constant so that no collisions occur (steady behavior).  
And at the lowest flow rate (regime c), the fluid periodically drips 
from the tank, rather than jets as with the higher flow rates, creating 
a regular spacing between perturbations near the tank outlet.  The long 
time between drips allows the film connecting consecutive perturbations to 
thin and subsequently become unstable to smaller capillary perturbations.  
(Figure~1 in \cite{kliak01} illustrates these three regimes of behavior.)  
Simulations of a Stokes flow model developed by KDB qualitatively captured 
two of the three observed behaviors (regimes b and c), while the behavior 
associated with the highest flow rate (regime a) could not be replicated 
\cite{kliak01}.

Craster \& Matar (CM) \cite{cras06} select a different scaling than KDB 
to derive an evolution equation for the free surface.  Using traveling 
wave solutions, their Stokes flow model quantitatively predicted the 
perturbation speed and height of regime a measured by KDB.  The model 
also qualitatively captured regime c, though the steady pattern of 
perturbation spacing found in regime b could not be matched with 
traveling wave solutions.  In experiments, CM observed regime b near 
the tank outlet, however, they found the regularly spaced pattern of 
perturbations disassembled itself further down the fiber.  From this 
observation, CM concluded regime b is a transient rather than steady 
regime \cite{cras06}.

The contradiction in observations of regime b (steady behavior) 
by KDB and CM motivated us to look more closely at the steady and 
unsteady states by examining the dynamics of the perturbations 
where they initially form along the fiber.  In experiments using 
fluids with different densities, surface tensions and viscosities 
we observe regimes a (unsteady), b (steady) and c (dripping) as 
described by KDB \cite{kliak01}, though the focus of this paper 
is on regimes a and b.  In our experiments, we observe the flow 
transitions abruptly from unsteady to steady behavior at a critical 
flow rate $Q_c$ (the value of $Q_c$ is dependent on the particular 
fluid) \cite{smolka06}.  In a recent independent study, Duprat 
{\it et al.} \cite{duprat07} explain the transition from regime a 
(unsteady behavior) to regime b (steady behavior) as a transition 
from a convective to absolute instability.  In their experiments
with silicone oil using a range of fiber and orifice radii, they 
find the transition occurs only at intermediate film thicknesses 
and for sufficiently small fiber radii; at thin or thick film 
thickness, they find the perturbation behavior remains convective 
(unsteady) \cite{duprat07}.  These criteria may explain why CM 
did not observe the steady dynamics of regime b in their experiments 
with silicone oil.  Here we find the transition from unsteady to 
steady behavior is also correlated to the rate at which perturbations 
naturally form along the fiber.  For $Q<Q_c$ (steady case), the rate 
of perturbation formation is constant.  As a result the position 
along the fiber where perturbations form is nearly fixed, and the 
spacing between consecutive perturbations remains constant as they 
travel 2~m down the fiber.  For $Q>Q_c$ (unsteady case), the rate 
of perturbation formation is modulated.  As a result the position 
along the fiber where perturbations form oscillates irregularly, and 
the initial speed and spacing between perturbations varies resulting 
in the coalescence of neighboring perturbations further down the fiber.

The paper is organized as follows.  The experimental setup and 
properties of the unperturbed flow are presented in Section~\ref{sec-exper}.  
Measurements of the perturbation growth are compared to analytical 
predictions for Stokes and moderate Reynolds number conditions in 
Section~\ref{sec-perform}.  The perturbation behavior exhibited 
in regimes a (unsteady) and b (steady) near the tank outlet is 
closely examined in Section~\ref{sec-stunst}.  Conclusions are 
provided in Section~\ref{sec-conc}.

\section{Experimental Setup \& Details}
\label{sec-exper}

\begin{figure}[h]	
\begin{center}	

\includegraphics [width=2 in,angle=90]{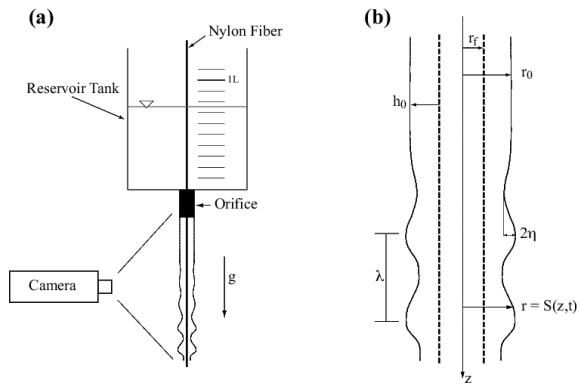}
															 
\caption{(a) Schematic of experimental setup. 
(b) Cross-section of an annular film flowing down the outside 
of a fiber of radius $r_f$ (not to scale). The unperturbed film 
radius measured from the fiber centerline is $r_0,$ unperturbed 
film thickness $h_0$, perturbation amplitude $\eta(z,t)$ and 
perturbation wavelength $\lambda$. }  
\label{f-jetsch}										
\end{center}					
\end{figure}												

\noindent
\subsection{Experimental Apparatus}

The experimental setup consists of viscous fluids, a reservoir 
tank/orifice assembly, nylon fishing line, a high-speed digital 
imaging camera, illumination, a computer and edge-detection 
software (a schematic of the experiment is shown in 
Figure~\ref{f-jetsch}(a)).  The reservoir tank (6~L capacity) 
is graduated at 100~mL increments to measure flow rate.  An 
orifice, machined with a flat edge (inner radius~=~0.11~cm, 
outer radius~=~0.16~cm, length~=~2.6~cm), is attached to the 
bottom of the tank to ensure a reproducible solid/fluid/air contact 
line in the experiment.  A nylon fiber ($r_f = 0.029$~cm) anchored 
from above, passes through the center of the tank/orifice assembly, 
and is held vertically plumb with weights attached 2~m below the 
orifice.  The fluid, which is gravitationally forced from the tank, 
coats the fiber to create an annular film.  To reduce air currents 
and other noise during data collection, the entire apparatus is 
enclosed by an aluminum frame with plastic sheet sidewalls and top.

The motion of the annular film is recorded using a high-speed digital 
imaging camera (Phantom v4.2) at rates between 1000-4000 frames/s and 
an image size of 64~x~512 pixels$^2$ with the camera focused on 
approximately the upper 10~cm of the fiber.  Illumination is obtained 
using silhouette photography following \cite{scor90}, with a 250~W lamp, 
an experimental grade one-way transparent mirror (Edmund Scientific,
A40,047) and high contrast reflective screen material (Scotchlite 3M 7615).  
Movies of the annular film are recorded and downloaded to a computer 
using camera software.  The free surface of the film is determined from 
movie images using an edge-detection algorithm.  The algorithm locates 
the free surface by interpolating the maxima positions in a gradient 
image; the gradient image is produced using the Frei-Chen operator 
\cite{frei77}.  The algorithm can detect the edge of the film free 
surface to within approximately 1/10$^\text{th}$ of a pixel which for 
the screen resolution in our experiments corresponds to $\approx 0.002$~cm.

The experimental fluids consist of castor oil, vegetable oil (Crisco)
and an 80:20 glycerol/water solution (by weight).  The temperature, 
density ($\rho$), surface tension ($\gamma$), dynamic viscosity 
($\eta$) of the fluids, and the framing rate and screen resolution
used in the experiments are listed in Table~\ref{tab-1}.  The surface 
tension was measured at room temperature using a Fisher 21 tensiomat and 
viscosity was measured using a temperature controlled cone and plate 
rheometer (Brookfield, Model DV-III+).  The fluid temperature varied 
by less than 4.6\%, 1.9\% and 1.4\% in the castor oil, vegetable oil 
and glycerol solution experimental runs, respectively.  We note that 
the selection of fluids allows us to independently probe the influence 
of surface tension or viscosity on flow behavior since castor oil and 
vegetable oil have comparable surface tension while vegetable oil and 
the glycerol solution have comparable viscosity.


\begin{table}
\caption{\label{tab-1}Fluid properties and experimental conditions.}
\begin{ruledtabular}
\begin{tabular}{ccccccc}
Fluid & Temp. & $\rho$ & $\gamma$ & $\mu$ & Framing & Screen \\
& & & & & Rate & Resolution\\
& $^o$C& $\frac{\text{g}}{\text{cm}^3}$ & $\frac{\text{dyn}}{\text{cm}}$ & 
$\frac{\text{g}}{\text{cm} \cdot \text{s}}$ & 
$\frac{\text{frames}}{\text{s}}$ 
& $\frac{\text{cm}}{\text{pixel}}$\\
\hline
 Castor Oil & 21.9 & 0.94 & 36.8 & 8.48 & 1000 & 0.0190 \\
 Vegetable Oil & 21.6 & 0.92& 34.3 & 0.58 & 4000 & 0.0180\\
 Glycerol Solution & 21.2 & 1.21 & 60.4 & 0.54 & 2000 & 0.0186\\
\end{tabular}
\end{ruledtabular}
\end{table}


\noindent
\subsection{Base Flow Properties}

\begin{figure}[h]	
\begin{center}	

\includegraphics [width=3 in]{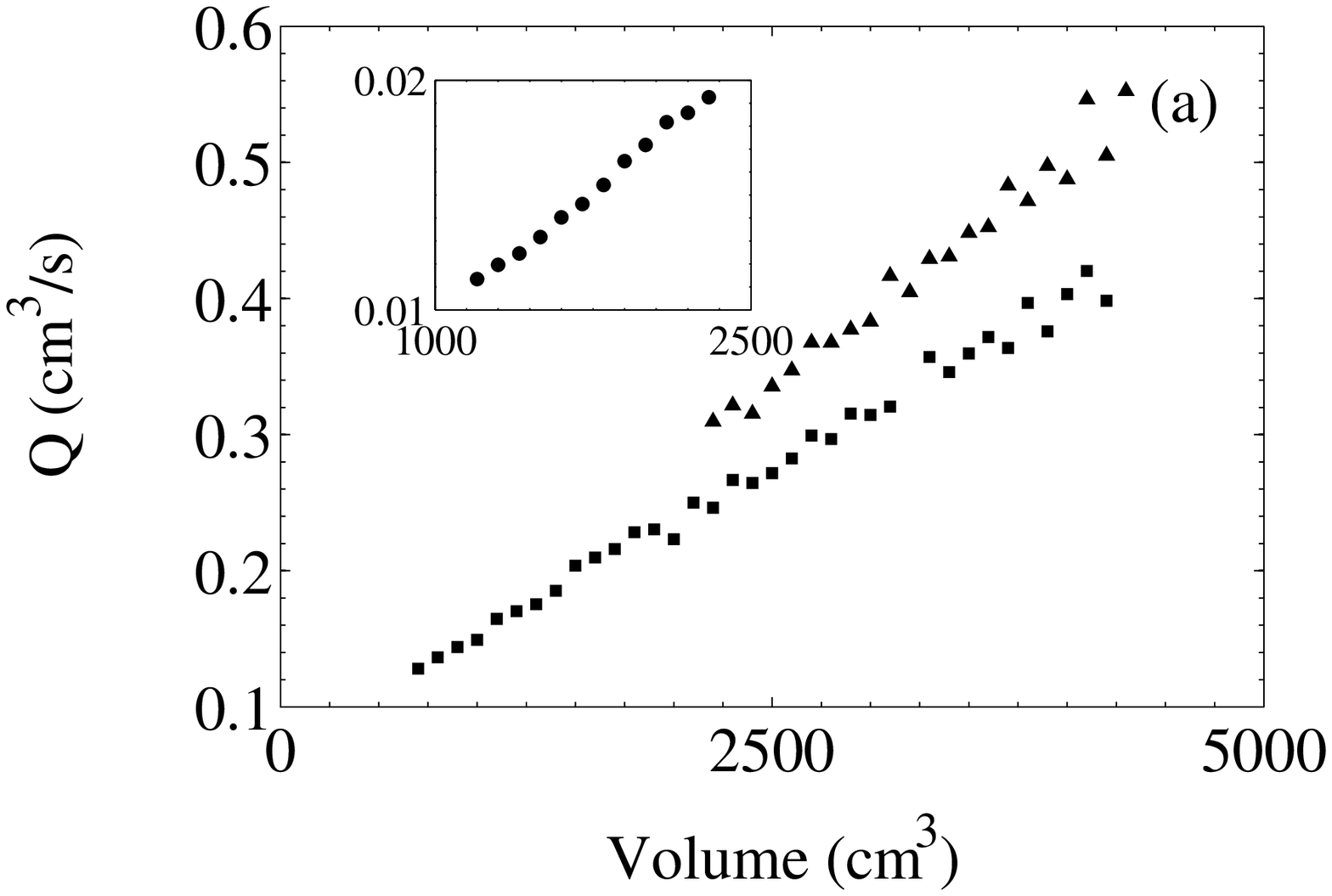}
\includegraphics [width=3 in]{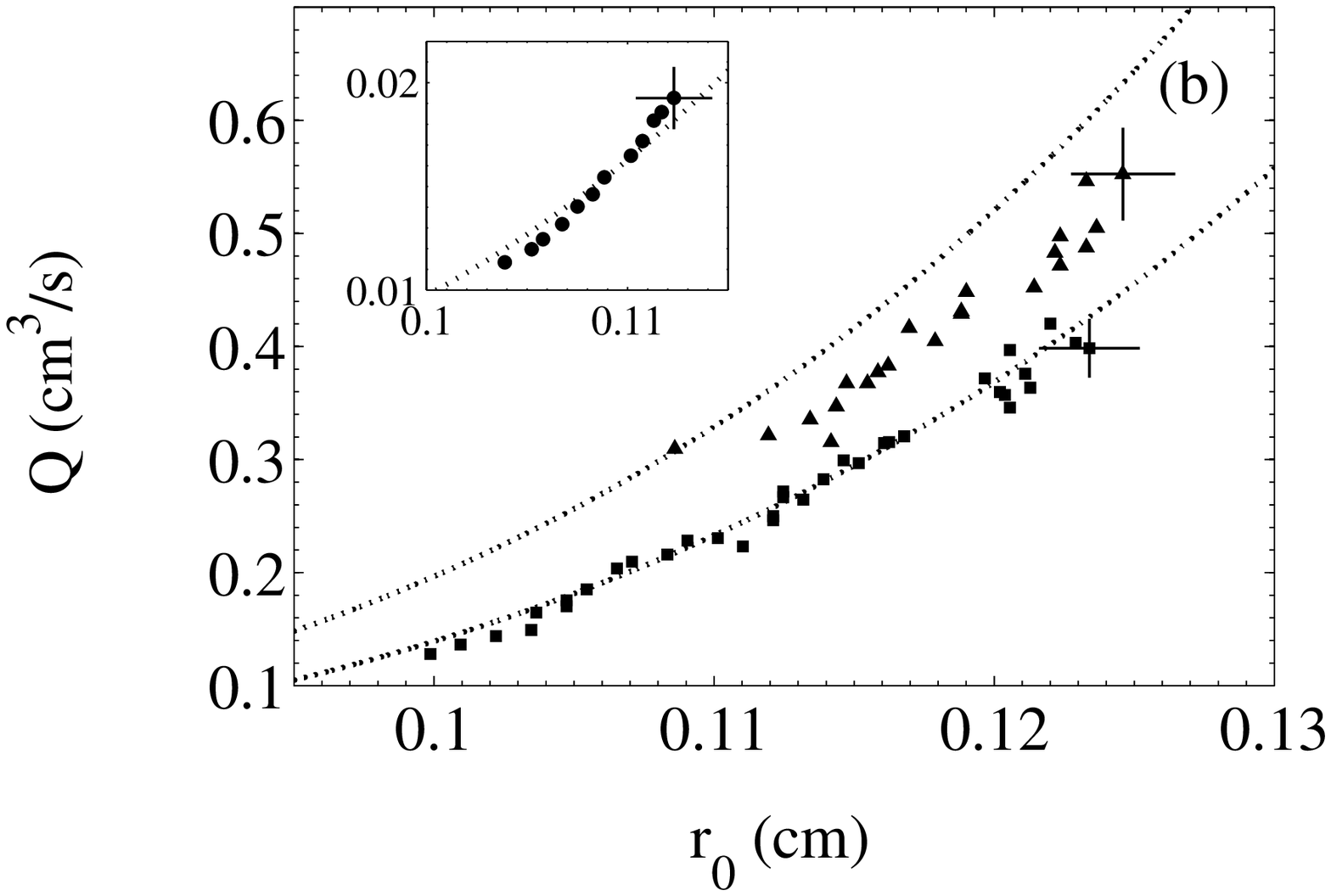}

\caption{(a) Experimental flow rate as a function of the 
reservoir tank volume.  (b) Flow rate as a function of the
unperturbed film radius: experimental data (symbols),
theory given by (\ref{e-qtheory}) with $r_f = 0.029$~cm (dotted); 
bars represent the resolution of experimental measurements.  
Symbols correspond to: vegetable oil (square),  glycerol solution 
(triangle) and castor oil (circle).}
\label{f-QvV}											
\end{center}				
\end{figure}

In each experiment, the reservoir tank drained under the influence of 
gravity and the elapsed time and tank volume were recorded as the fluid 
passed each 100~mL mark to determine the flow rate ($Q$).  Data was 
collected only while the flow was jetting from the orifice and an 
unperturbed region of the film free surface was present near the orifice 
(corresponding to regimes a and b \cite{kliak01}).  Measurements of the 
flow rate during each experimental run for vegetable oil (square), 
glycerol solution (triangle) and castor oil (circle), is shown in 
Figure~\ref{f-QvV}(a).  (Note that in each experiment the flow rate decreases 
as the tank volume decreases.)  Figure~\ref{f-QvV}(a) shows 
that the flow rate increases linearly over the range of tank volume 
used in each experimental run.  Furthermore, the flow rate decreases 
with increasing viscosity ($Q_{castor}$ is an order of magnitude less 
than $Q_{glycerol}$ and $Q_{vegetable}$), and increases with increasing 
surface tension ($Q_{glycerol} > Q_{vegetable}$).

The unperturbed film radius $r_0$, measured from the fiber centerline 
to the unperturbed free surface (see Figure~\ref{f-jetsch}(b)), was 
measured using edge-detection software.  Figure~\ref{f-QvV}(b) shows
the unperturbed radius increases with flow rate in each experimental 
run.  This trend is similar to behavior observed in dip coating in 
which the film thickness increases with the velocity at which the 
fiber is drawn from the fluid source at sufficiently low withdrawal 
rates \cite{quere99,ruck02,deryck96,shen02}.  Comparing data for 
the glycerol solution (triangle) and vegetable oil (square) experiments, 
we find that at a fixed flow rate higher surface tension (glycerol 
solution) results in a thinner unperturbed film.  In our experiments, 
the range of the ratio of unperturbed film thickness to fiber radius 
is $\,2.5 < h_0/r_f < 3.3,\,$ which means the films are thick.

The relevant forces that characterize the dynamics of an annular film can 
be determined by considering two dimensionless groups, the Reynolds number 
and Bond number.  Following CM \cite{cras06}, we define characteristic 
length and velocity scales as 
\begin{equation}
\mathcal{L} =\gamma/(\rho g r_0), \qquad
\mathcal{V}~=~\rho g r_0^2/\mu,
\end{equation}
where $g$ is the gravitational constant of acceleration, the Reynolds 
number as $\text{Re} = \rho \mathcal{V L}/ \mu,$ which compares inertial 
to viscous effects, and the Bond number as $\text{Bo} = \rho g r_0^2/\gamma,$ 
which compares gravitational to surface tension effects \cite{cras06}.  
The values of $\text{Re}$ and $\text{Bo}$ for our experiments, listed in 
Table~\ref{tab-2}, indicate inertial effects are negligible for castor oil 
$(\text{Re} \sim O(10^{-2})),$  in contrast to the vegetable oil and 
glycerol solution experiments $(\text{Re} \sim O(10)),$ and surface 
tension dominates over gravitational effects in the formation of 
interfacial perturbations for all three fluids $(\text{Bo} \sim O(10^{-1})).$


\begin{table}
\caption{\label{tab-2}The Reynolds number and Bond number measured in 
experiments.}
\begin{ruledtabular}
\begin{tabular}{ccc}
Fluid & $\text{Re}$ & $\text{Bo}$ \\
\hline
 Castor Oil & 0.049 - 0.054 & 0.26 - 0.32  \\
 Vegetable Oil & 9.4 - 11.6 & 0.26 - 0.40 \\
 Glycerol Solution & 27.2 - 31.2 & 0.23 - 0.30  \\
\end{tabular}
\end{ruledtabular}
\end{table}


Experimental \cite{cras06,kliak01} and theoretical 
\cite{cras06,kliak01,sisoev06,trif92} studies often assume the base
flow of an annular film is steady, unidirectional and parallel.  
Under these conditions, the unperturbed flow with free surface located at 
$r=r_0$ and constant pressure field $p(r,z,t) = p_0$ is described by the 
boundary value problem for the axial velocity $w(r)$
\begin{equation}
\mu r^{-1}\partial_r(r \partial_r w) + \rho g =0, \quad
w(r_f) = 0, \quad \partial_r w(r_0) = 0,\label{e-bvp}
\end{equation}
where the boundary conditions include no-slip at the fiber and zero 
tangential stress at the free surface.  Eqs.~(\ref{e-bvp}) can be solved 
exactly for the axial velocity to obtain
\begin{equation}
w(r) = {\rho g \over 4\mu} \left[2r_0^2 \ln(r/r_f) 
+ r_f^2 -r^2\right].\label{e-uniflow}
\end{equation}
Using (\ref{e-uniflow}), the flow rate of an annular film can be 
expressed in terms of $r_f$ and $r_0$ 
\begin{eqnarray}
Q(r_f, r_0) &=& 2 \pi \int_{r_f}^{r_0} r w(r)\, dr \nonumber\\
&=& {\rho g \pi \over 8 \mu}\left(4r_f^2r_0^2 
+ 4 r_0^4\ln(r_0/r_f) - r_f^4 - 3r_0^4\right).
\label{e-qtheory}
\end{eqnarray}
By comparing (\ref{e-qtheory}) to experimental data, we can test 
the assumption that the base flow of an annular film is a steady, 
unidirectional parallel flow.

Figure~\ref{f-QvV}(b) shows a comparison of the flow rate as a 
function of the unperturbed film radius for vegetable oil, glycerol 
solution and castor oil measured directly in experiments (symbols) 
and using (\ref{e-qtheory}) with $r_f = 0.029$~cm (dotted line).  
We find excellent agreement between (\ref{e-qtheory}) and the 
experimental data in the castor oil (circle) and vegetable oil 
(square) experiments which indicates the flow in these experiments 
is well approximated by a unidirectional parallel flow.  The theory 
overestimates $Q$ by as much as 12\% in the glycerol solution 
experiment (triangle).  This is not surprising since the assumptions 
on the flow are equivalent to steady Stokes flow.  With 
$\text{Re} \approx 30,$ we suspect the Stokes flow assumption is not 
valid for the glycerol solution experiment.  Based on the comparison
in Figure~\ref{f-QvV}(b), we estimate (\ref{e-uniflow}) and 
(\ref{e-qtheory}) are valid when $\text{Re} \le 10.$  The 
experiments conducted by KDB \cite{kliak01} and CM \cite{cras06} 
meet this criterion, thus we conclude their assumption that the flow 
is steady, unidirectional and parallel is indeed valid.

In the proceeding sections, we examine the dynamics of a perturbed 
annular film including the initial formation and longer-time dynamics
of interfacial perturbations along the free surface.

\section{Perturbation Growth}
\label{sec-perform}

The image in Figure~\ref{f-anjet} illustrates the capillary 
instability an annular viscous film undergoes as the unperturbed
free surface becomes unstable to undulations that develop into 
large amplitude perturbations.  Next we present experimental 
observations on the growth of these interfacial perturbations 
and compare their initial growth to theoretical predictions developed
from linear stability analysis.  Before proceeding, we first 
recount relevant stability results developed in the Stokes 
\cite{cras06} and moderate Reynolds number flow \cite{sisoev06} limits.

\subsection{Linear Stability Results}
\label{sec-theory}

\subsubsection{Stokes Flow}

Craster \& Matar \cite{cras06} derive a long-wave Stokes flow evolution 
equation for the free surface of the annular film, $r=S(z,t),$ under the 
assumption that the unperturbed film radius $r_0$ is small relative to 
the capillary length $l_c = \gamma/\rho g r_0$ 
(i.e., $\text{Bo} = r_0/l_c  = \rho g r_0^2/\gamma \ll 1$) and the 
Reynolds number is sufficiently small ($\text{Re} \le O(1)$), to obtain
\begin{widetext}
\begin{eqnarray}
\p_{\hat{t}} (\hat{S}^2) 
+ {1 \over 8} \p_{\hat{z}} \left[\left[\p_{\hat{z}} \left({1 \over \hat{S}} 
- (\text{Bo})^2 \p_{\hat{z}\hat{z}} \hat{S}\right) - 1\right]
\times (\alpha^4 -4 \hat{S}^2 \alpha^2 
+3\hat{S}^4 -4\hat{S}^4 \ln(\hat{S}/\alpha))\right] = 0,\label{e-cras}
\end{eqnarray}
\end{widetext}
where $\hat{S}, \hat{z}, \hat{t}$ are dimensionless variables 
satisfying the scalings
\begin{equation}
S=r_0\hat{S}, \qquad z = \mathcal{L}\hat{z}, \qquad 
t=\mathcal{L}\hat{t}/\mathcal{V},
\end{equation}
and $\alpha = r_f/r_0.$
Conducting a linear stability analysis by perturbing about the base flow 
\begin{equation}
\hat{S}(\hat{z},\hat{t})=1 + \hat{S}_1 e^{(im\hat{z}+\sigma \hat{t})},
\label{e-perth}
\end{equation}
where $m$ is the (real) dimensionless wavenumber and $\sigma$ is the 
(complex) dimensionless growth rate CM obtain the following dispersion 
relation for the growth rate
\begin{eqnarray}
\sigma = {m^2 \over 16}(\text{Bo}^2m^2 -1)(\alpha^4-4\alpha^2 +3 
+4 \ln(\alpha))\nonumber\\
 - {im \over 2}\left(\alpha^2-1 -2 \ln(\alpha)\right). 
\label{e-stdisp}
\end{eqnarray}

\subsubsection{Moderate Reynolds Number Flow}

In an analytical study, Trifonov \cite{trif92} derived model equations
for fluid flowing down the inside or outside of a vertical cylinder at 
moderate Reynolds number; the model includes evolution equations for 
the film thickness, $h(z,t) = S(z,t)-r_f,$ and volumetric flow rate, 
$q(z,t)$.  In a recent study, Sisoev {\it at al.} \cite{sisoev06} 
rescale Trifonov's equations for flow down the outside of a vertical
cylinder, casting the model in terms of a generalized falling film
model \cite{shka04} (see Eqs.~(11)-(13) in \cite{sisoev06}).  
Conducting a linear stability analysis of the rescaled equations by 
perturbing about the base solution 
\begin{subequations}
\begin{eqnarray}
\hat{h}(\hat{z},\hat{t}) &=& 1 + \hat{h}_1 e^{i(m\hat{z}-\sigma \hat{t})},\\
\hat{q}(\hat{z},\hat{t}) &=& 1 + \hat{q}_1 e^{i(m\hat{z}-\sigma \hat{t})},
\label{e-hq}
\end{eqnarray}
\end{subequations}
where $m$ is the (real) dimensionless wavenumber and $\sigma$ is the 
(complex) dimensionless growth rate, Sisoev {\it et al.} obtain a 
dispersion relation for $\sigma$ satisfying
\begin{equation}
\sigma^2 + (i a_{1,0} - a_{1,1}m)\sigma 
+ {m \over 1 + \eps}(-a_{0,3}m^3 - a_{0,1}m + i a_{0,0})=0,\label{e-redisp}
\end{equation}
where $\eps = h_0/r_f$ and the constant coefficients 
$a_{1,0}, a_{1,1}, a_{0,3}, a_{0,1}, a_{0,0}$ are defined in the 
Appendix.  The variables $\hat{h}, \hat{z}, \hat{t}$ are dimensionless 
quantities satisfying the scalings
\begin{equation}
h = h_0\hat{h}, \qquad z = {h_0 \over \kappa}\hat{z}, \qquad 
t = {h_0 \over \kappa U}\hat{t},
\end{equation}
where 
\begin{equation}
\kappa = \left({\rho g h_0^2 \over\gamma}\right)^{1/3}, \qquad 
U = {Q \over 2\pi r_f h_0},
\end{equation}
represent a stretching parameter and characteristic velocity
scale, respectively.  We note that the long-wave model used by
Sisoev {\it et al.} was derived under the assumption that 
$\kappa^2 \ll 1.$

The stability results developed by Craster \& Matar \cite{cras06} 
and  Sisoev {\it at al.} \cite{sisoev06} are temporal analysis 
(since $m \in {\mathbb R}$ and $\sigma \in {\mathbb C}$), and thus 
model the case in which interfacial perturbations grow in amplitude
everywhere along the film \cite{keller73}.  Our interest is in 
testing these stability predictions by comparing the theoretical 
dispersion relations (\ref{e-stdisp}) and (\ref{e-redisp}) to the 
growth rate of perturbations measured in experiments conducted in 
the Stokes and moderate Reynolds number flow limits.

\subsection{Experimental Observations of Perturbation Formation}


\begin{figure}[h]	
\begin{center}	

\includegraphics[width=2.7 in,angle=90]{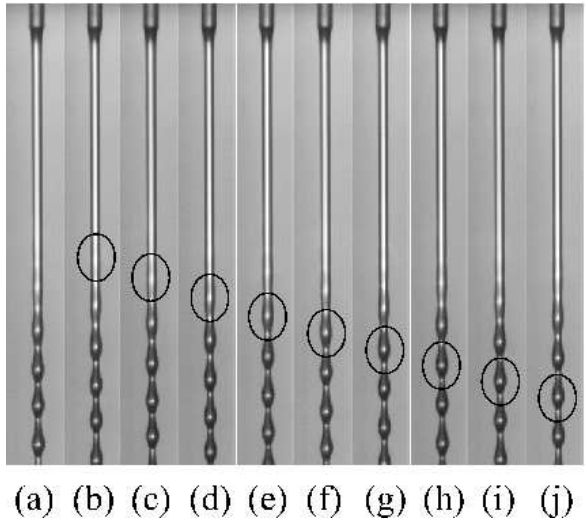}													 
\caption{An annular film of castor oil flowing down the outside of a 
thin nylon fiber.  The film loses uniformity approximately 5.4~cm 
from the orifice indicated by the circled area in frame (b).  
Subsequent frames track the position and growth of this nascent 
perturbation.  Time between images is 0.42~s and elapsed time~=~3.7~s. 
Image height~=~9.7~cm.}
 
\label{f-cascoll}											
\end{center}
						
\end{figure}												

Figure~\ref{f-cascoll} shows a series of images tracking the
formation of a perturbation along an annular film of castor oil.
In frame~(b), a small amplitude perturbation first appears along 
the film approximately 5.4~cm from the orifice (circled region).  
Frames~(c)-(j) track the position of this perturbation as it grows 
in amplitude and saturates in size.  Once formed, the perturbation 
continues moving down the fiber (not shown).  Since the perturbation 
grows in amplitude as it travels down the fiber, the flow is spatially 
(convectively) unstable rather than temporally (absolutely) unstable 
to perturbations \cite{keller73}.

\begin{figure}[h]	
\begin{center}	

\includegraphics[width=3.2 in]{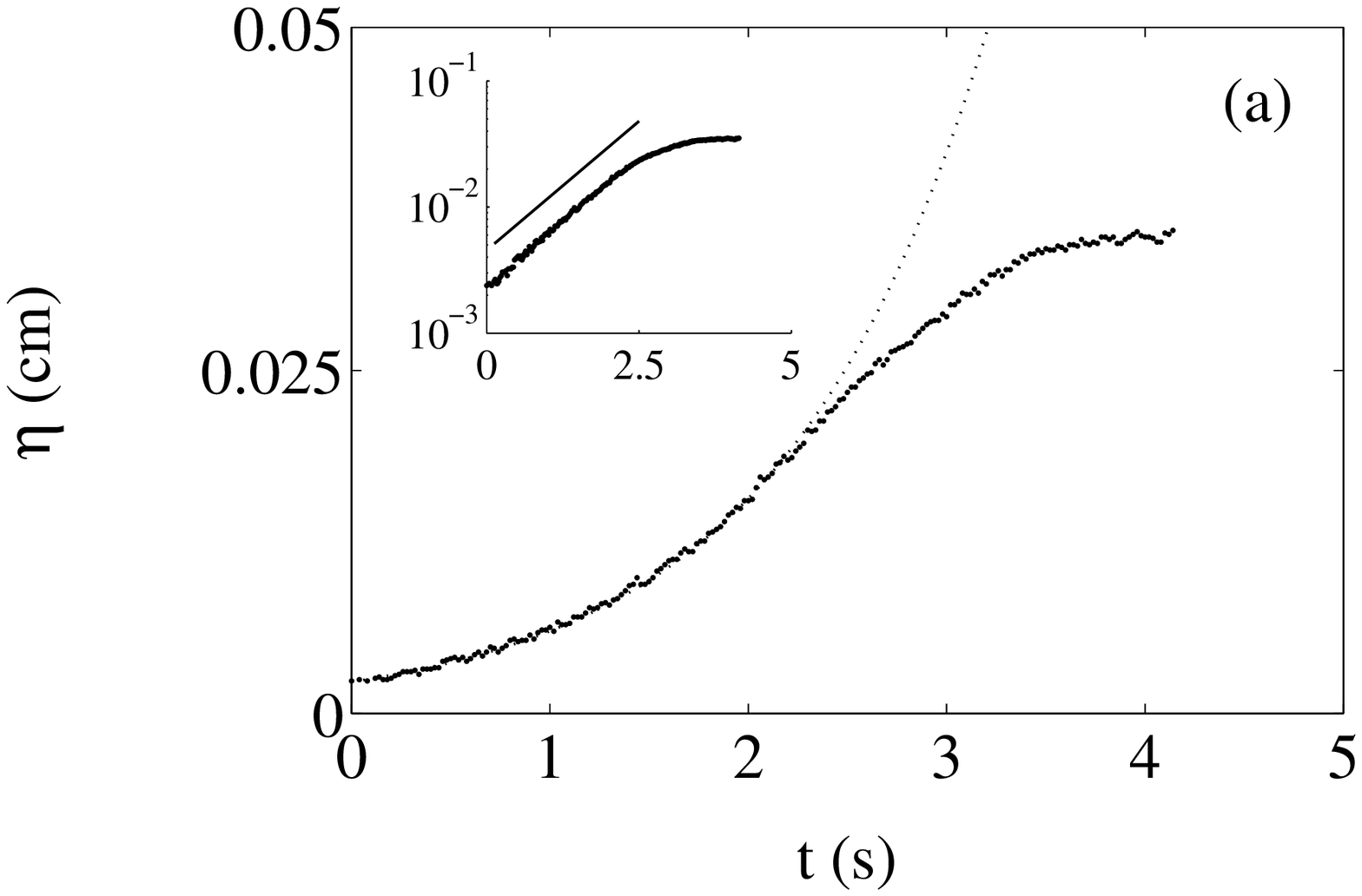}
\includegraphics[width=3 in]{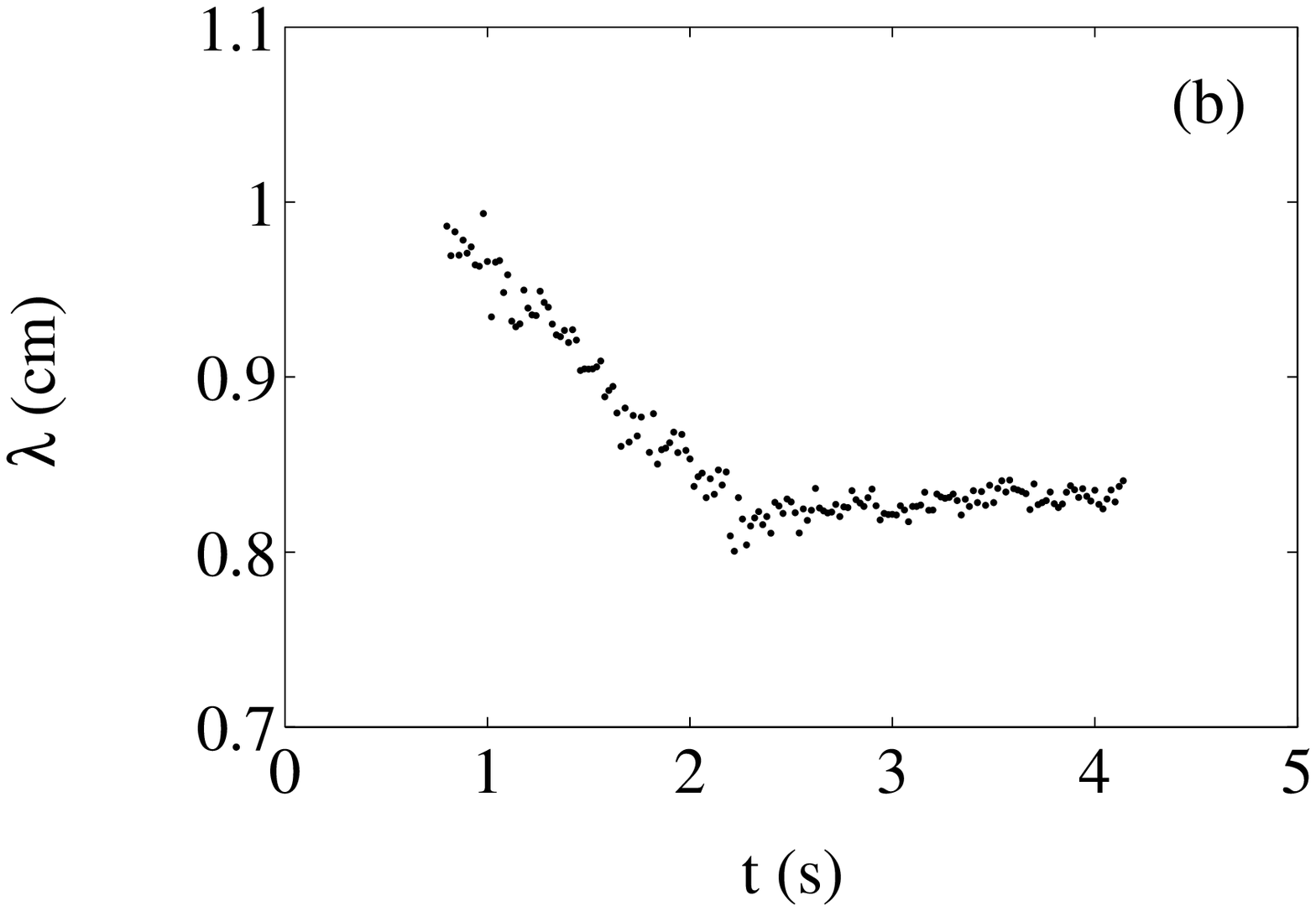}															 
\caption{(a) Amplitude ($\eta$) and (b) wavelength ($\lambda$) 
of a nascent perturbation as a function of time for castor oil 
($Q=0.0194$~cm$^3$/s); experimental data corresponds to the perturbation 
tracked in Figure~\ref{f-cascoll}. (a) The initial amplitude growth 
of the perturbation is exponential (shown in inset on log-linear scale) 
followed by nonlinear saturation. Dotted line: fit of data to 
$\eta (t) = 0.0023 e^{0.959t}$ corresponding to growth between frames 
(b)-(g) in Figure~\ref{f-cascoll}.  (b)  The wavelength decreases 
during the time interval that the amplitude grows exponentially and 
then saturates in length as the amplitude saturates in size.} 

\label{f-gr}
									
\end{center}						
\end{figure}												

To characterize the growth of a perturbation we measure the amplitude
$\eta$ (half the radial distance from first minima to first maxima) and 
the wavelength $\lambda$ (the axial distance from first to second maxima)
as shown in Figure~\ref{f-jetsch}(b) using edge-detection software; both 
measurements are made in the moving reference frame of the 
perturbation.  The data shown in Figure~\ref{f-gr} corresponds to the 
perturbation followed in Figure~\ref{f-cascoll}.  Figure~\ref{f-gr}(a) 
shows the nascent growth of the amplitude is exponential (inset) followed 
by a slower phase as the perturbation saturates in size 
($\eta_{saturate} =0.035$~cm).  The growth rate for the initial formation 
of the amplitude is determined from a least squares fit of the data to 
an exponential function yielding the dimensional growth rate 
$\sigma_{dim} = 0.959$~s$^{-1}$ (fit indicated by dotted line in 
Figure~\ref{f-gr}(a)).  The wavelength of this perturbation decreases 
from $\lambda =0.98$~cm to 0.80~cm during the time interval that the 
amplitude grows exponentially $(0 \le t \le 2.34~\text{s})$, before 
saturating in length to $\lambda = 0.83$~cm, as shown in 
Figure~\ref{f-gr}(b).  The decrease in $\lambda$ during the exponential 
phase of growth indicates the annular film is unstable to a range of 
wavenumbers ($=2\pi/\lambda),$ rather than to one fixed value.

The behavior displayed in Figure~\ref{f-gr} for the amplitude and 
wavelength is typical of observations made in the castor oil, 
vegetable oil and glycerol solution experiments.   Comparison of
perturbation growth to the Stokes (\ref{e-stdisp}) or moderate 
Reynolds number (\ref{e-redisp}) dispersion relations depend on 
the flow conditions in each experiment, specifically on the Reynolds 
and Bond numbers (provided in Table~\ref{tab-2}).  In the castor 
oil experiments, $\text{Bo} \sim O(10^{-1})$ and 
$\text{Re} \sim O(10^{-2}),$ thus satisfying the requirements of 
the Stokes model ($\text{Bo} \ll 1,$ $\text{Re} \le O(1)$).  Since 
$\text{Re} \approx 30$ in the glycerol solution experiments,
inertial effects cannot be ignored, and so we compare this
case to the moderate Reynolds number model.  With
$\text{Bo} \sim O(10^{-1})$ and $9.4 < \text{Re} < 11.6,$ the
vegetable oil experiments are on the border of the requirements
for the Stokes model.  In this case we compare the experimental 
data to both the Stokes and moderate Reynolds number dispersion 
relations.

\begin{figure}[h]	
\begin{center}	
	 
\includegraphics[width=3 in]{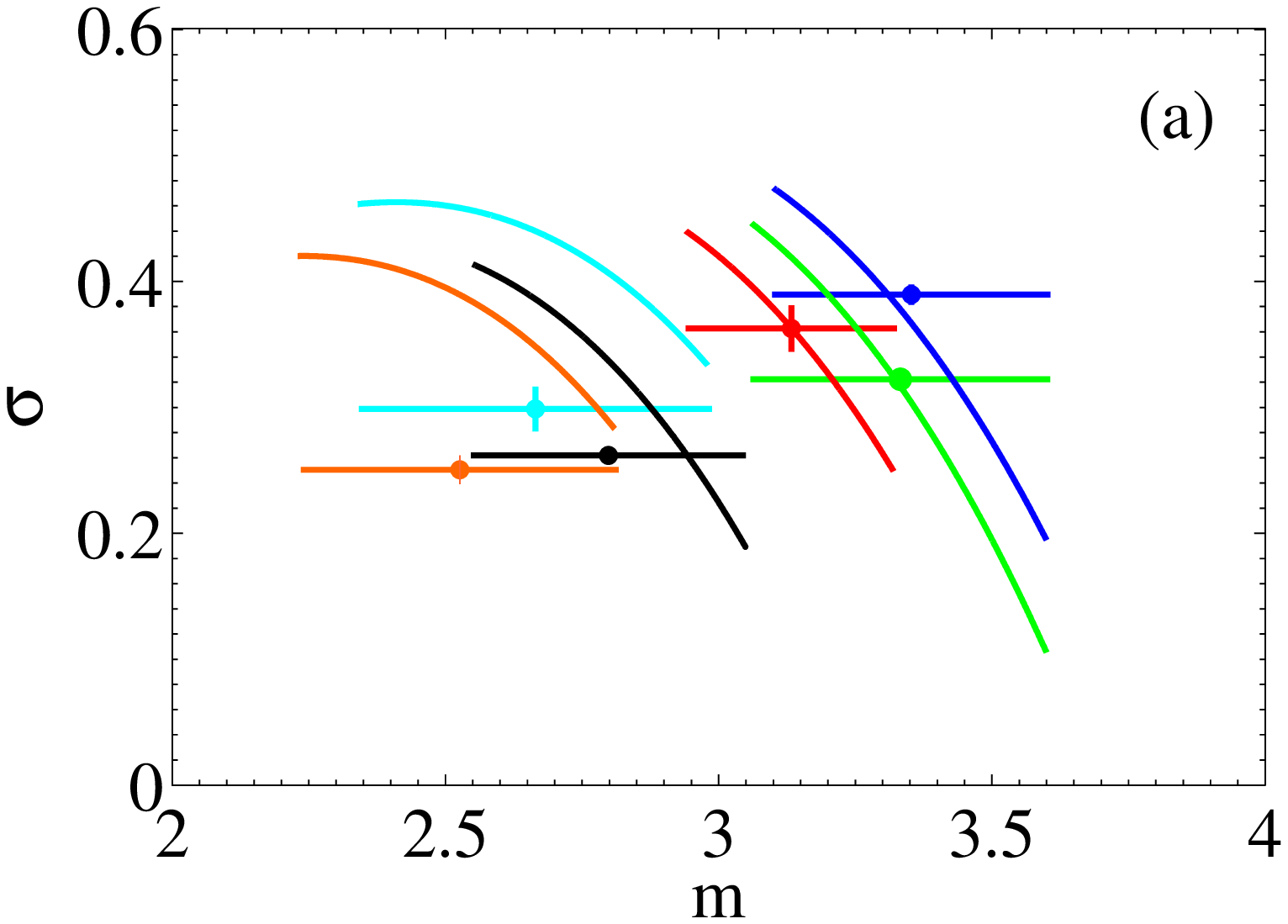}
\includegraphics[width=3 in]{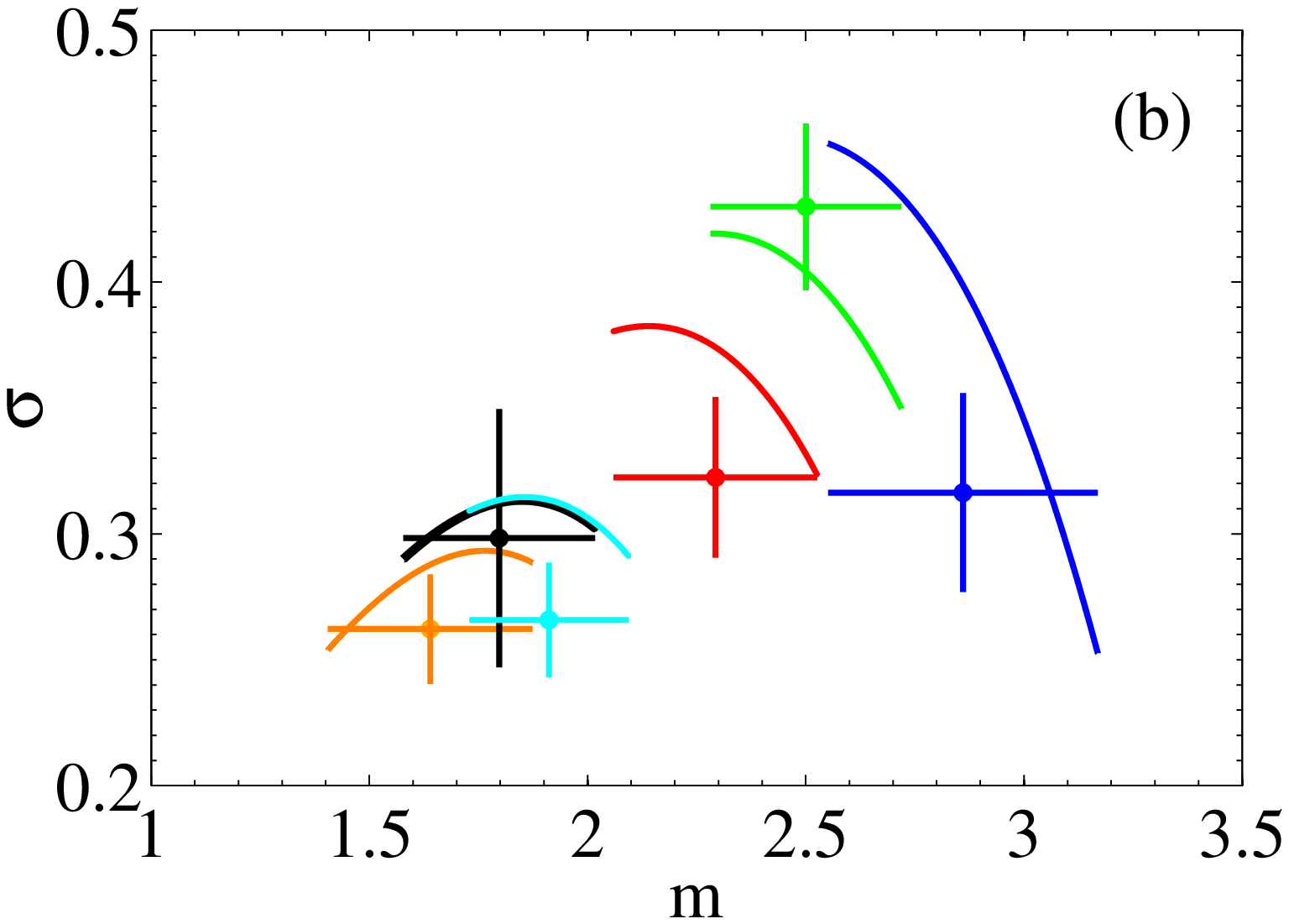}

\caption{(Color online) Dimensionless growth rate as a function of 
dimensionless wavenumber for experiments with (a) castor oil and (b) 
vegetable oil. Circles correspond to experimental data of the average
growth rate and wavenumber measured over several perturbations at 
(a): Q = 0.0194 - orange, 0.0171 - black, 0.0148 - cyan, 
0.0123 - red, 0.0111 - green and 0.0103 - blue (cm$^3$/s); (b): 
Q = 0.416 - orange, 0.383 - black, 0.342 - cyan, 
0.252 - red, 0.211 - green and 0.170 - blue (cm$^3$/s).  
Vertical bars represent the standard deviation of $\sigma$ and 
horizontal bars represent the range of $m$ measured during 
the period of exponential growth over all perturbations measured.  
Corresponding colored curves represent theoretical prediction
given by the Stokes flow dispersion relation (\ref{e-stdisp}) of 
${\mathcal Re}(\sigma)$ plotted over the range of $m$ measured 
in experiments \cite{cras06}.} 	

\label{f-CMdisp}								

\end{center}							
\end{figure}												

Figure~\ref{f-CMdisp} shows a comparison of the measured amplitude 
growth rate to the dispersion relation developed by Craster \& Matar 
in the Stokes flow limit (\ref{e-stdisp}) for (a) castor oil and 
(b) vegetable oil at various flow rates.  At a given flow rate, the 
growth rate for several perturbations was measured (8-12 perturbations 
for castor oil, and 15-44 for vegetable oil).  The average
dimensionless growth rate ($\sigma = \sigma_{dim}\mathcal{L}/\mathcal{V}$) 
and dimensionless wavenumber ($m=2\pi\mathcal{L}/\lambda$) measured 
over all the perturbations is denoted by a circle with each color 
corresponding to a different flow rate.  The vertical bars represent 
the standard deviation of all the growth rates measured at a given 
flow rate during the period of exponential growth.  Since the
perturbation wavelength decreases over a range of values during the 
exponential phase of growth, we cannot assign a single wavenumber to
its growth.  Instead, the horizontal bars represent the range of 
wavenumber measured during the period of exponential growth of all 
the perturbations.  The corresponding colored curves represent the 
real part of the growth rate predicted by 
(\ref{e-stdisp}) plotted over the range of wavenumber measured at each 
flow rate.  We consider the theory to be in agreement with the experimental 
data (at a given flow rate) if the theoretical curve overlaps 
the rectangular region defined by the resolution bars of the data.  
Figure~\ref{f-CMdisp} shows that the Stokes theory is 
in agreement with four of the six castor oil experiments and with five 
of the six vegetable oil experiments.  In the other three experiments, 
the theory overestimates the measured values by 10 to 13\,\%. The 
quantitative agreement between theory and experimental data is excellent, 
a significant result considering: (i) the comparison is between a temporal 
stability theory and a spatial instability of the film,  and (ii) the value 
of the Reynolds number in the vegetable oil experiments, 
$\text{Re} \approx 10,$ is slightly higher than the criteria for the 
Stokes theory, $\text{Re} \le O(1).$

\begin{figure}[h]	
\begin{center}	

\includegraphics[width=3 in]{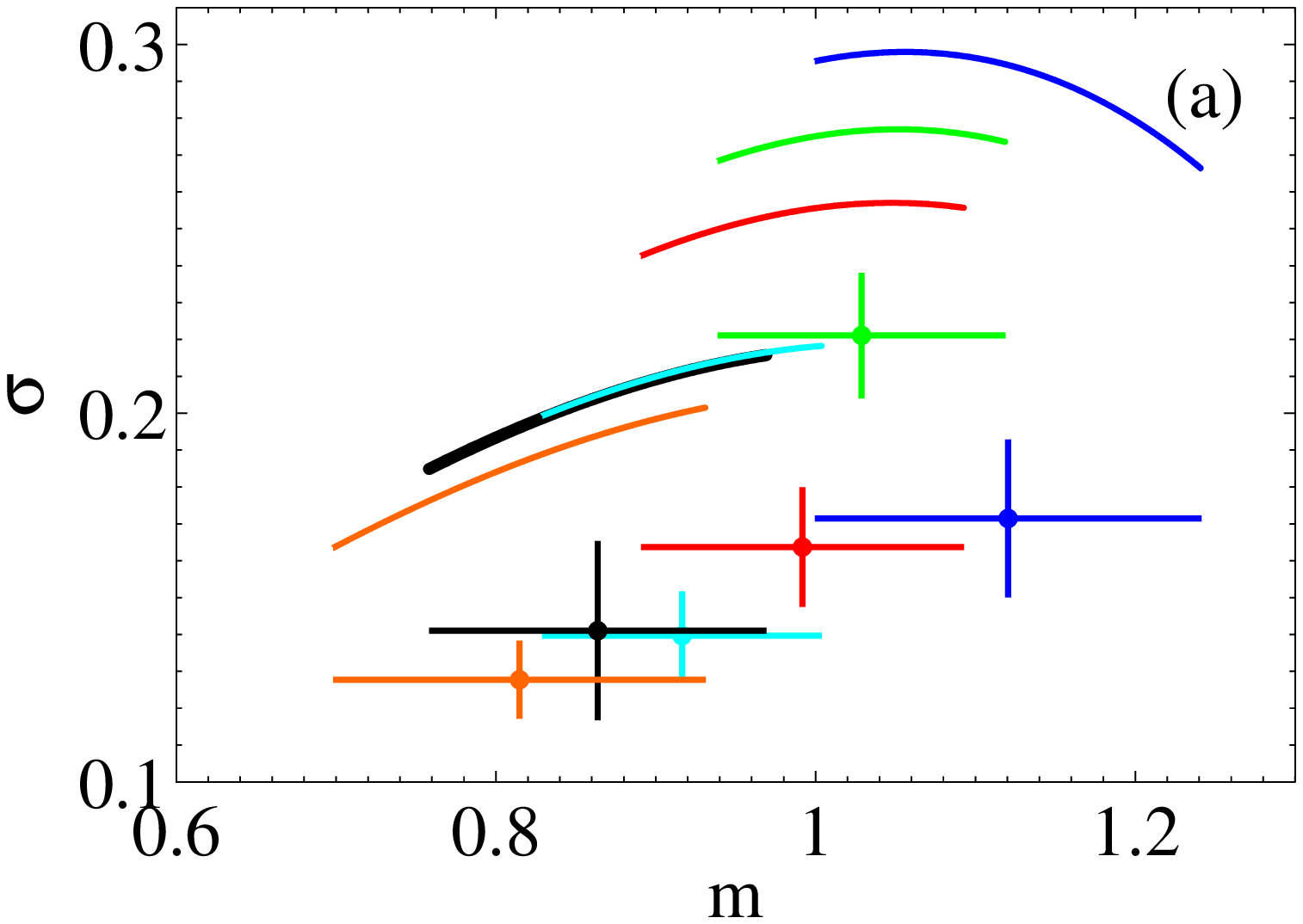}
\includegraphics[width=3.1 in]{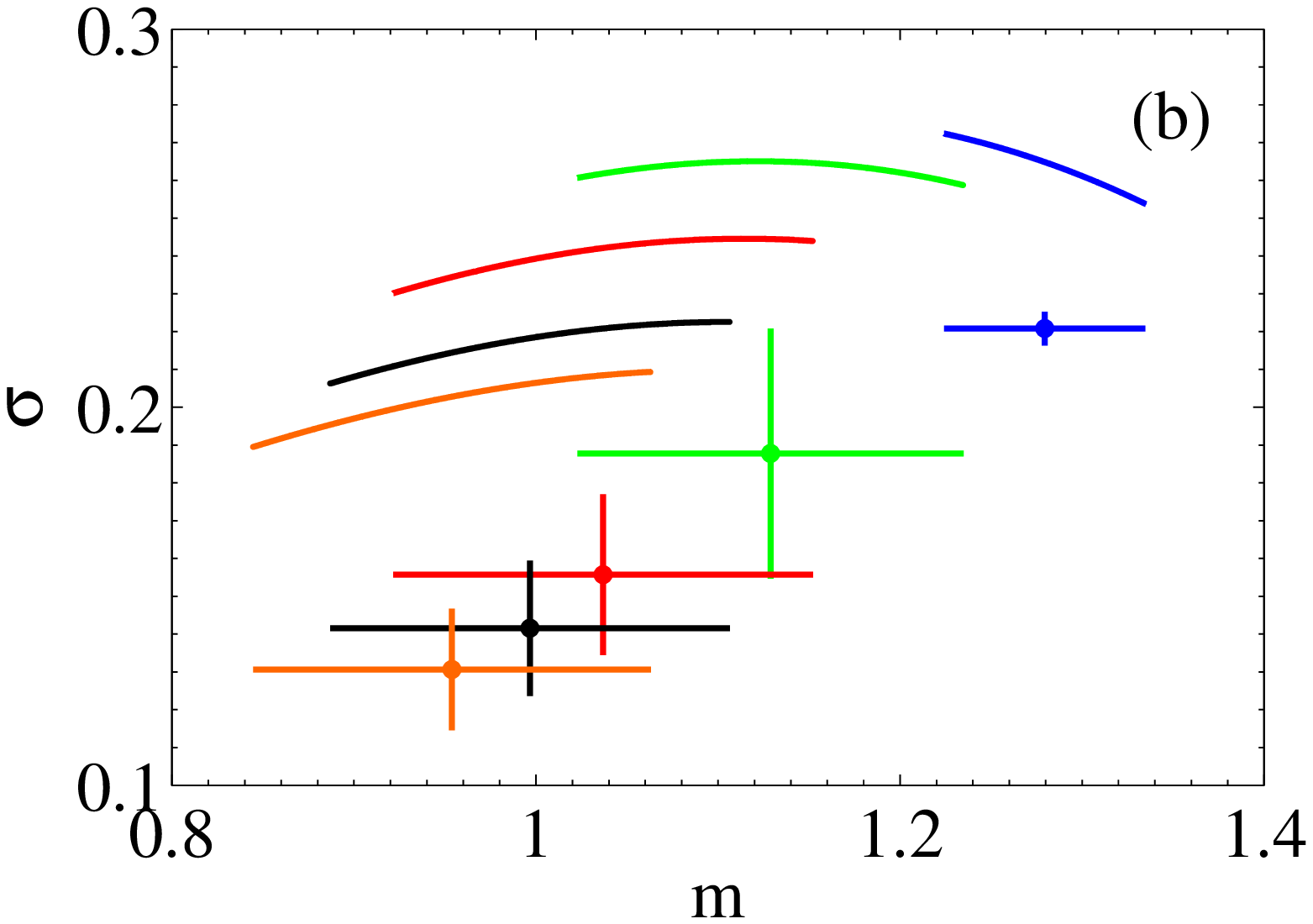}

\caption{(Color online) Dimensionless growth rate as a function of 
dimensionless wavenumber for experiments with (a) vegetable oil and 
(b) glycerol solution. Circles correspond to experimental data of the 
average growth rate and wavenumber measured over several perturbations 
at (a): Q = 0.416  - orange, 0.383 - black, 0.342 - cyan, 
0.252 - red, 0.211 - green and 0.170 - blue (cm$^3$/s); (b): 
Q = 0.538 - orange, 0.493 - black, 0.437 - red, 0.381 - green 
and 0.325 - blue (cm$^3$/s).  Vertical bars represent the standard 
deviation of $\sigma$ and horizontal bars represent the range of 
$m$ measured during the period of exponential growth over all 
perturbations measured.  Corresponding colored curves represent 
theoretical prediction given by the moderate Reynolds number flow 
dispersion relation (\ref{e-redisp}) of ${\mathcal Im}(\sigma)$ plotted 
over the range of $m$ measured in experiments \cite{sisoev06}.} 												 
\label{f-Sdisp}											
\end{center}				
\end{figure}												

Figure~\ref{f-Sdisp} shows a comparison of the measured amplitude growth 
rate to the dispersion relation developed by Sisoev {\it et al.} in the 
moderate Reynolds number limit (\ref{e-redisp}) for (a) vegetable oil and 
(b) glycerol solution at various flow rates.  The data and theory are 
presented in a similar fashion to Figure~\ref{f-CMdisp} with the exception 
that the dimensionless growth rate and wavenumber are given by 
$\sigma = \sigma_{dim}h_0/(\kappa U)$ and $m=2\pi h_0/(\kappa \lambda)$, 
and the growth rates for the glycerol solution experiments are averaged 
over 88 to 102 perturbations.  We recall that the moderate Reynolds number
model is valid as long as $\kappa^2 \ll 1$; in all of the experiments
shown in Figure~\ref{f-Sdisp}, $\kappa^2 \sim O(10^{-1}).$  We find the 
moderate Reynolds number model overestimates the measured growth rates 
by 28 to 50\,\% in the vegetable oil experiments and 15 to 48\,\% in the 
glycerol solution experiments, as shown in Figure~\ref{f-Sdisp}.  Clearly, 
the Stokes model is more accurate at predicting the growth rate of the 
perturbations in the vegetable oil experiments than the moderate Reynolds 
number model.  This is somewhat surprising since the vegetable oil 
experiments slightly exceed the Reynolds number limit of the Stokes 
model, but satisfy the assumption on $\kappa^2$ for the moderate 
Reynolds number model.  While the theoretical growth rates in 
Figure~\ref{f-Sdisp} are on the same order of magnitude as the measured 
values, the quantitative match between theory and data is not strong.

We note that the range of the measured amplitude growth rate in 
experiments (indicated by the vertical bars in Figures~\ref{f-CMdisp} 
\& \ref{f-Sdisp}) varies by fluid and flow rate.  For castor oil 
(Figure~\ref{f-CMdisp}(a)), the range is fairly small which we 
attribute to the low Reynolds number ($\text{Re} \sim O(10^{-2})$) in 
the experiments.  For the experiments with vegetable oil 
(Figures~\ref{f-CMdisp}(b) and \ref{f-Sdisp}(a)) and glycerol solution 
(Figure~\ref{f-Sdisp}(b)) with $Q > 0.345$~cm$^3$/s, the range of the
measured amplitude growth rates is large.  Naively, one could attribute 
this to the higher Reynolds number in these experiments 
($10 \lesssim \text{Re} \lesssim 30$).  This is, however, not the complete 
picture.  Notice the range is significantly smaller for the glycerol solution 
experiment at $Q=0.325$~cm$^3$/s (blue (rightmost) data set in 
Figure~\ref{f-Sdisp}(b)).  The difference in this data set compared to the 
other glycerol solution and vegetable oil sets is in the behavior of the 
perturbations.  The perturbation behavior in the sets with a large range of 
amplitude growth rates is unsteady, while the behavior in the blue glycerol 
solution data set is steady.  (The notion of unsteady and steady perturbation 
behavior will be explained in detail in Section~\ref{sec-stunst}.)  
Therefore, we find the range of amplitude growth rate of the perturbations 
is correlated to both the Reynolds number of the flow and the longer-time 
dynamics of the perturbations.

Next, we examine the dynamics of perturbations after their initial 
formation and explain a physical mechanism that controls a known 
transition in the flow from unsteady to steady perturbation behavior.

\section{Steady and Unsteady Perturbation Dynamics}
\label{sec-stunst}

The dynamics of interfacial perturbations along an annular film flowing
down a vertical fiber can be broken down into three essential stages: 
(i) initial exponential growth of the perturbation amplitude accompanied
by a decrease in wavelength; (ii) nonlinear saturation of the perturbation 
amplitude and wavelength; and (iii) longer-time behavior in which the 
perturbation wavelength may (unsteady - see Figure~\ref{f-anjet}) or may 
not (steady) vary along the fiber; this last stage has been noted in other 
experimental studies \cite{cras06,duprat07,kliak01}.  Here we explain a 
physical mechanism that controls this third stage of dynamics.

\begin{figure}[h]	
\begin{center}	
\includegraphics[width=2.7 in,angle=90]{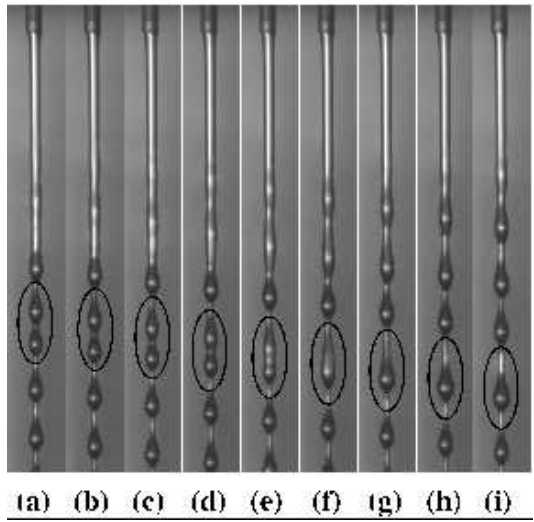}

\caption{Coalescence of two perturbations along an annular film of glycerol 
solution. $Q~=~0.359~$cm$^3$/s, time between images is 0.0125~s, elapsed 
time = 0.1~s, image height = 9.7~cm.} 												 
\label{f-coal}
									
\end{center}						
\end{figure}												

\begin{figure*}	
\begin{center}	

\includegraphics[width=2.75 in,angle=90]{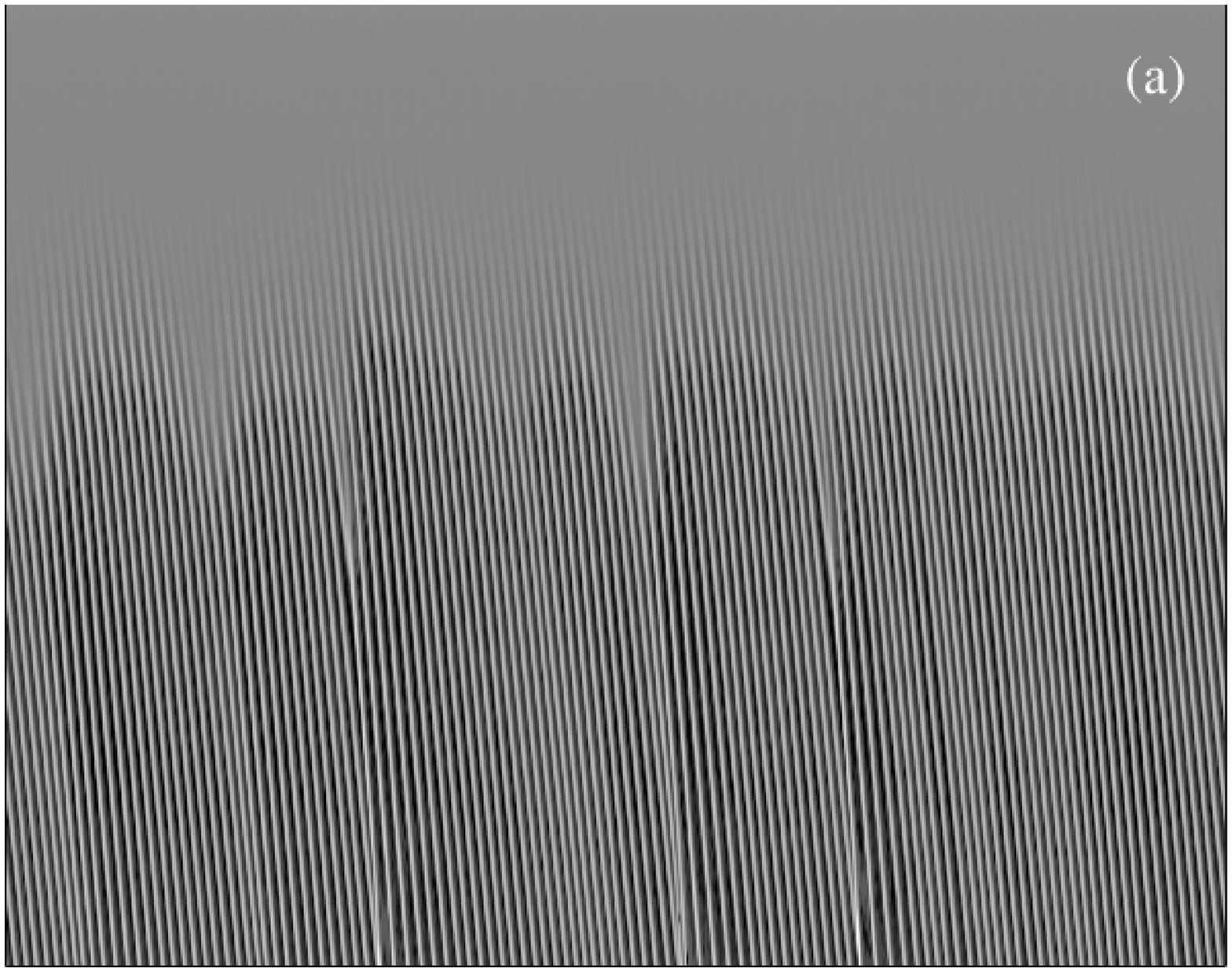}
\includegraphics[width=2.75 in,angle=90]{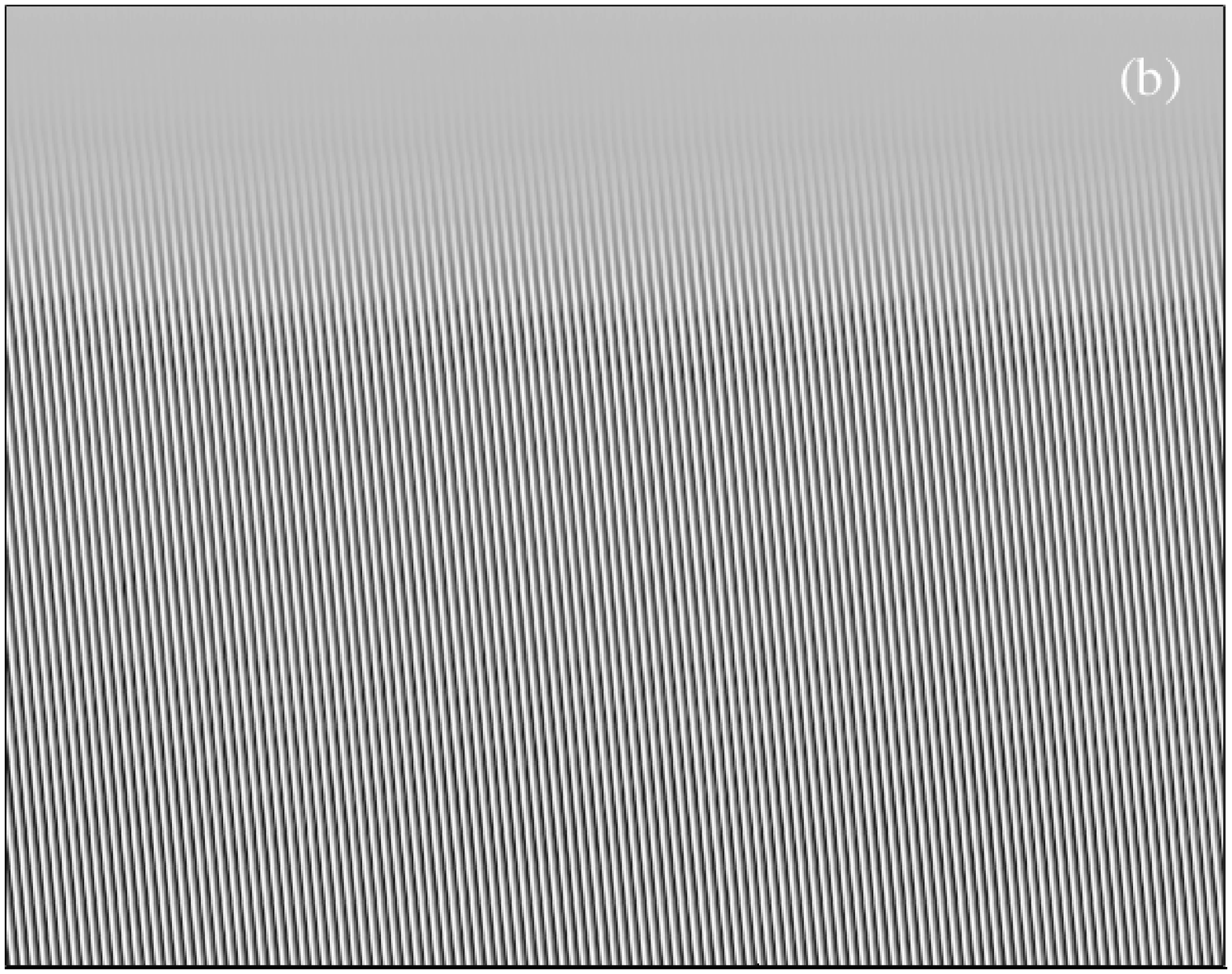}

\caption{Space-time plot illustrating the motion of perturbations along 
a vertical fiber during (a) unsteady and (b) steady behavior.  
Lighter (darker) regions correspond to thicker (thinner) regions of the 
fluid interface with light characteristic lines indicating the location 
of perturbations along the fiber as a function of time.  Experimental 
fluid: glycerol solution ($Q_c = 0.345$~cm$^3$/s), (a) 
$Q = 0.347~$cm$^3$/s, (b) $Q = 0.336~$cm$^3$/s, elapsed time = 8.09~s, image 
height = 8.22~cm.  The top of each image is 0.58~cm below the orifice.} 												 
\label{f-streak}										
\end{center}					
\end{figure*}												

In experiments with all three fluids, we observe the perturbation
motion abruptly transitions from unsteady (regime a) to 
steady (regime b) behavior at a critical flow rate, $Q_c$  
($Q_c =$ 0.0095~cm$^3$/s for castor oil, 0.119~cm$^3$/s for 
vegetable oil and 0.345~cm$^3$/s for glycerol solution) \cite{smolka06}, 
similar to observations made by Duprat {\it et al.} in their experiments
with silicone oil \cite{duprat07}.  Following KDB
\cite{kliak01}, we define the flow to be steady if no perturbations 
coalesce as they travel down the full length of the fiber ($\approx$~2~m), 
and unsteady otherwise while the flow is jetting from the orifice.  An 
example of unsteady behavior in which two perturbations coalesce is shown 
in Figure~\ref{f-coal}.  Note: we will not be examining the dripping state, 
regime c \cite{kliak01}, which occurs at a lower flow rate, 
$Q_{drip} < Q_c$.  We find the transition from unsteady ($Q > Q_c$) 
to steady ($Q_{drip} < Q < Q_c$) behavior is robust in the sense that 
once an experiment transitions to steady behavior (as the flow rate 
decreases) it does not revert back to the unsteady state.

\begin{figure}[h]	
\begin{center}	

\includegraphics[width=3.5 in]{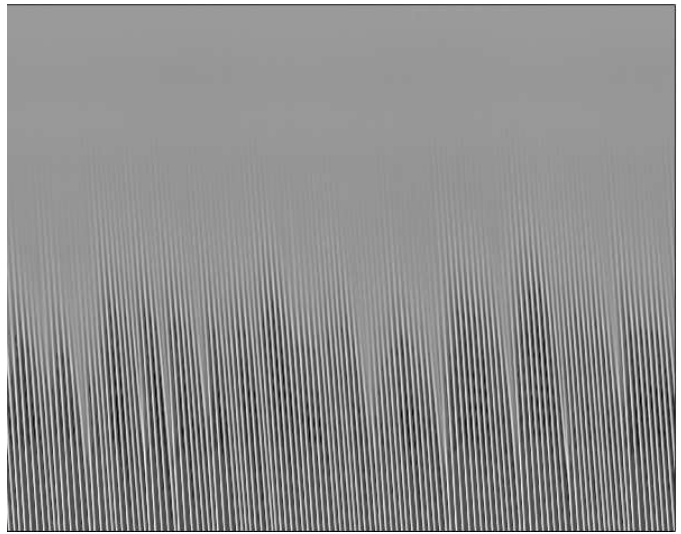}

\caption{Space-time plot illustrating the motion of perturbations along 
a vertical fiber during unsteady behavior.  Experimental fluid: glycerol 
solution; $Q = 0.516~$cm$^3$/s, elapsed time = 8.09~s, image height = 8.22~cm.  
The top of the image is 0.58~cm below the orifice.} 												 
\label{f-boundosc}										
\end{center}					
\end{figure}												

The space-time plots in Figure~\ref{f-streak} illustrate (a) unsteady 
and (b) steady perturbation behavior for experiments with  
glycerol solution.  Each plot is focused on 8.22~cm of the film, with 
the top of each plot located 0.58~cm below the orifice; the time span 
of each plot is 8.09~s.  The plots are created by mapping the radius 
of the free surface of the film, $r=S(z,t),$ to a gray level with lighter 
(darker) gray level corresponding to thicker (thinner) regions of the 
free surface.  The light characteristic lines in the plots indicate 
the location of perturbations as they move down the fiber, and their 
slope represent the speed of the perturbations.  Two features in the 
space-time plots distinguish the unsteady and steady perturbation behavior. 
First, the location along the fiber that perturbations form (which we 
refer to as the boundary) oscillates irregularly in the unsteady case 
and appears nearly fixed in the steady case \cite{smolka06}.  Second, 
in the unsteady case perturbations coalesce as faster moving perturbations 
collide into slower moving perturbations (indicated by intersecting 
characteristic lines), whereas in the steady case perturbations do not 
coalesce as they travel with the same terminal speed down the fiber 
(indicated by parallel characteristic lines) \cite{kliak01,smolka06}.  
The longer-time motion of the perturbations appears to be correlated 
to the motion of the boundary.  Notice in Figure~\ref{f-streak}(a) 
that large spatial variations in the boundary modulate the perturbation 
speed (i.e., the slope of the characteristic lines) which results 
in coalescence events later down the fiber.  In the steady case, there 
is no spatial variation in the boundary, and as a result, the perturbations 
remain equally spaced as they travel with constant terminal speed down the 
full length of the fiber (not shown in Figure~\ref{f-streak}(b)).  Our 
observations of the steady case (regime b) are consistent with those of 
KDB \cite{kliak01}.  Given the robustness of the 
steady dynamics in all of our experiments, we conclude this is not a 
transient state as CM report \cite{cras06}.  Finally, we note that 
when the flow is unsteady, the oscillation frequency of the boundary 
increases with increasing flow rate; for example, compare the boundary
frequency in Figures~\ref{f-streak}(a) and \ref{f-boundosc}.

\begin{figure}[h]	
\begin{center}	

\includegraphics[width=3.05 in]{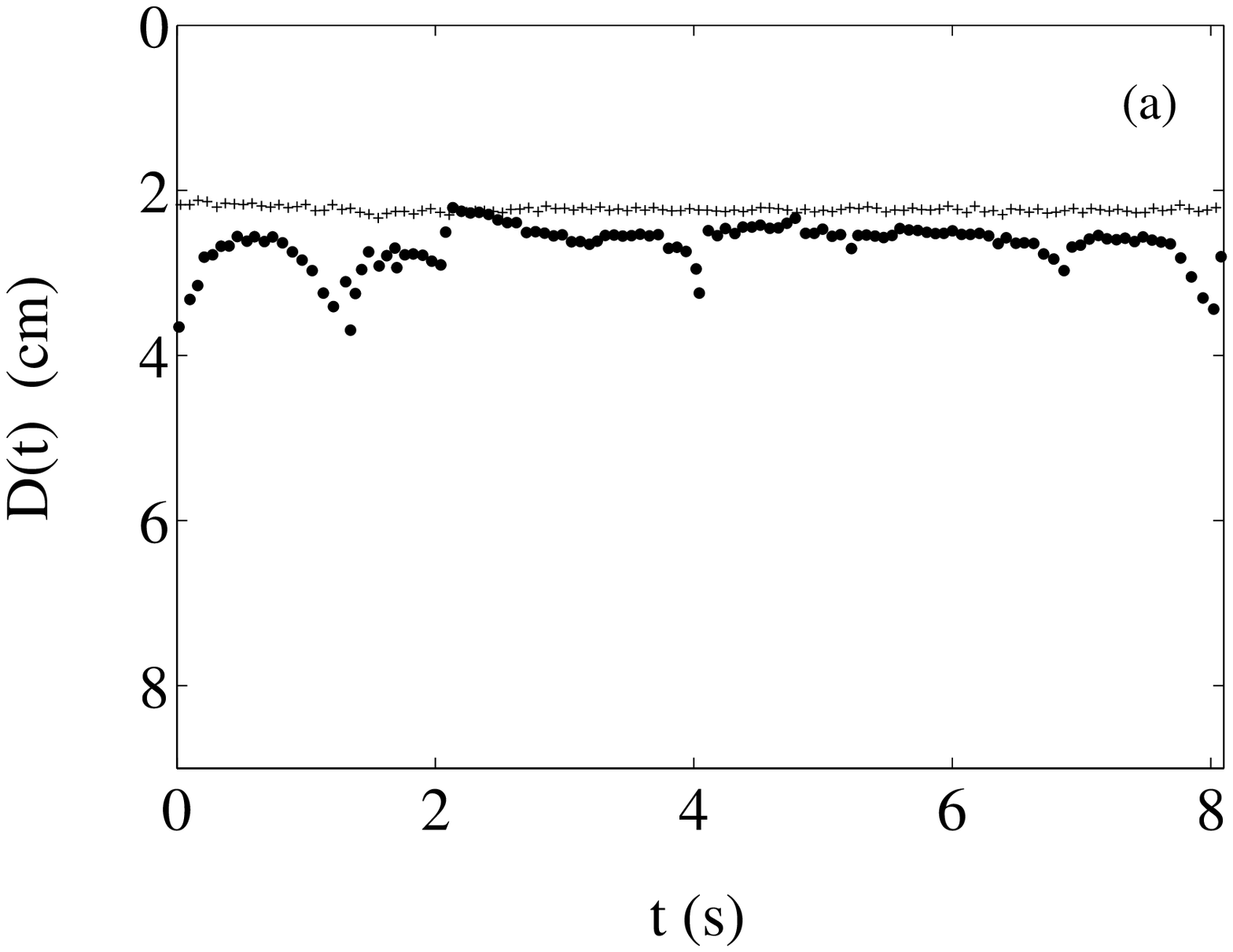}
\includegraphics[width=3 in]{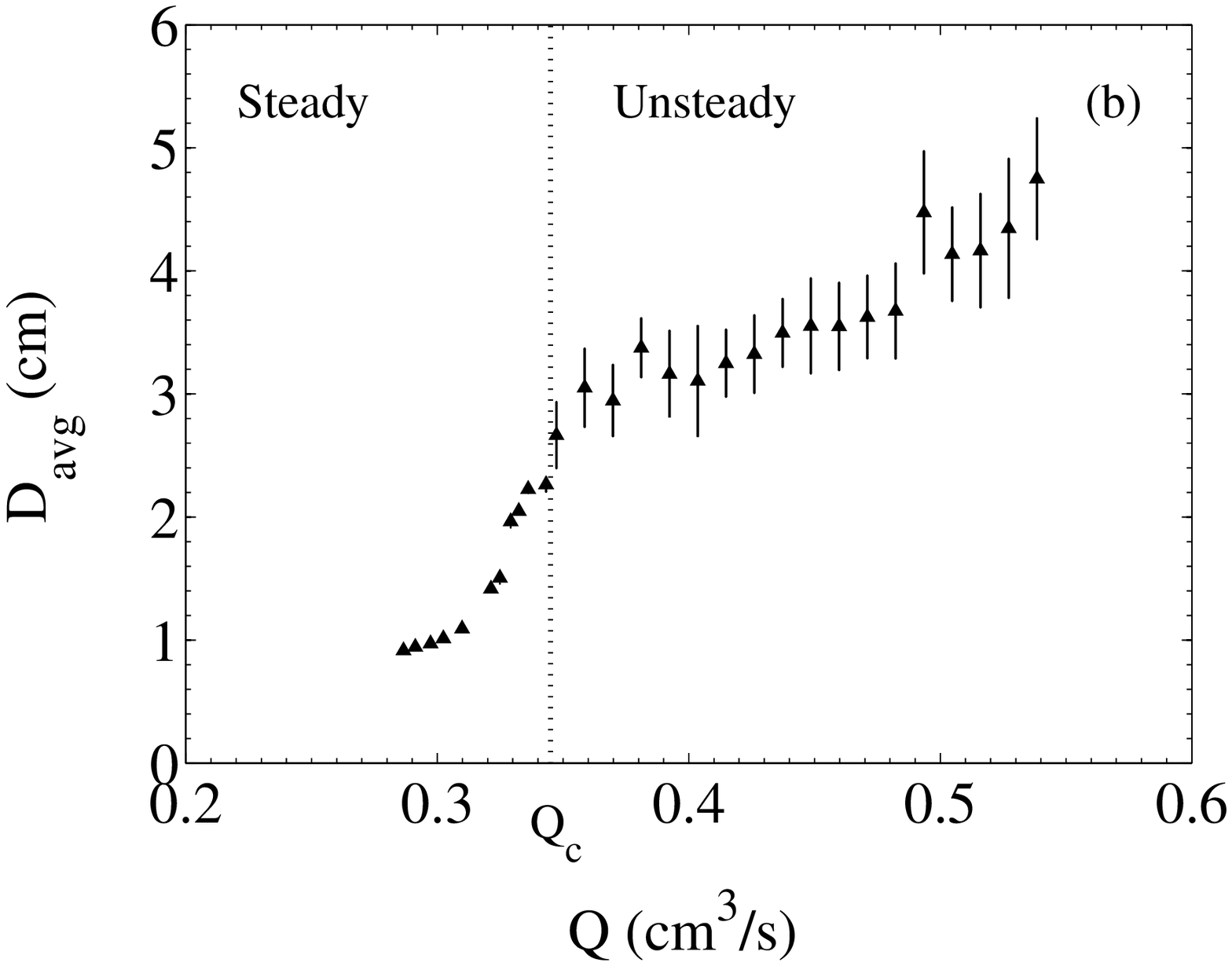}

\caption{(a) The distance from the orifice ($D$) where perturbations 
initially form along the fiber (i.e., the location of the boundary) as 
a function of time, corresponding to the data shown in 
Figure~\ref{f-streak}(a) for the unsteady case at $Q = 0.347~$cm$^3$/s 
($\bullet$) and in Figure~\ref{f-streak}(b) for the steady case at
$Q = 0.336~$cm$^3$/s (+). (b) Average distance ($D_\text{avg}$) from 
the orifice that perturbations form as a function of flow rate $Q$ 
for experiments with glycerol solution.  Vertical bars represent the 
standard deviation of $D$ over all the perturbations measured at a 
given flow rate.  The dotted vertical line denotes the transition flow 
rate separating steady and unsteady perturbation behavior.}									
			   
\label{f-Dvt}
									
\end{center}						
\end{figure}												

\begin{figure}[h]	
\begin{center}	

\includegraphics[width=3 in]{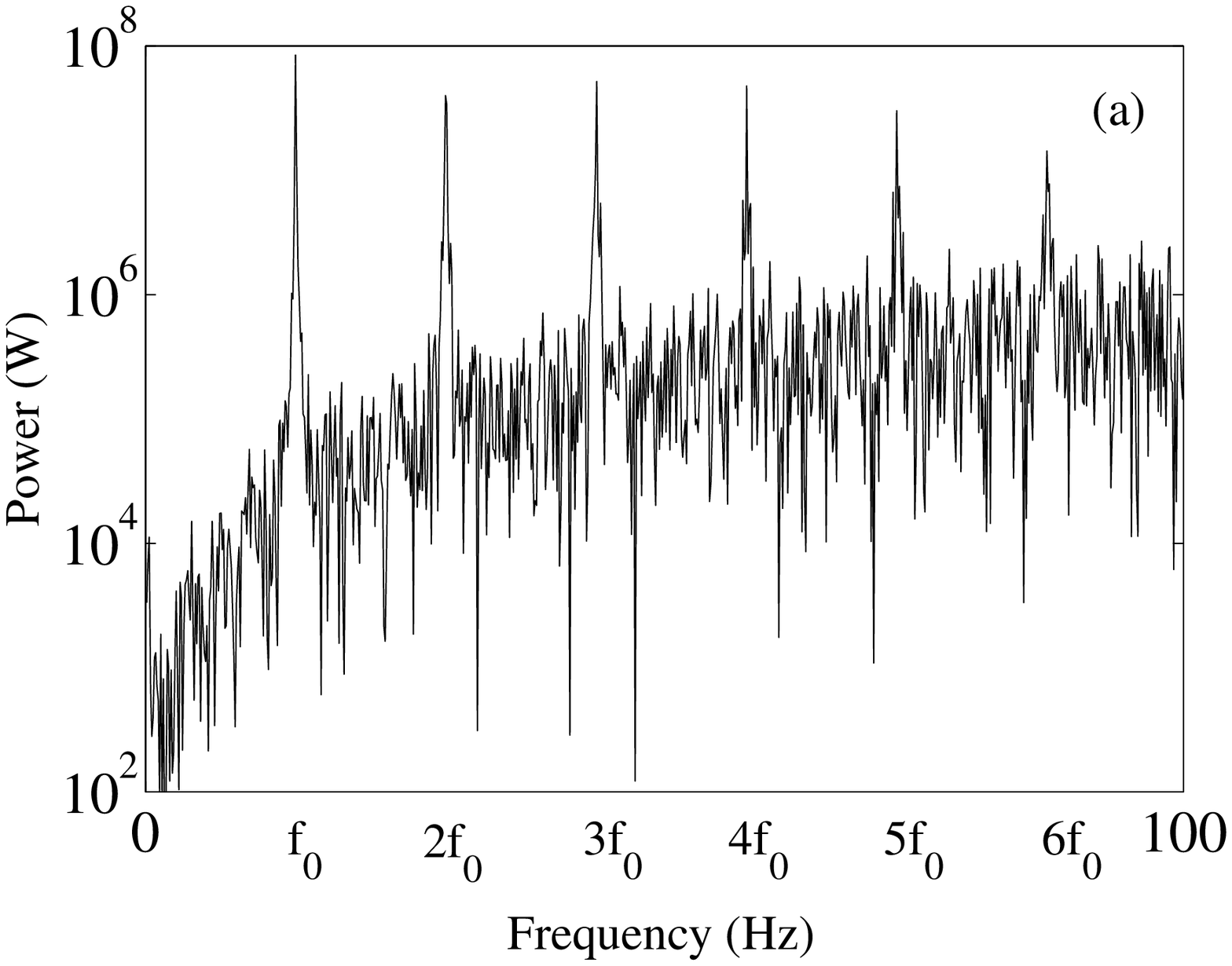}
\includegraphics[width=3 in]{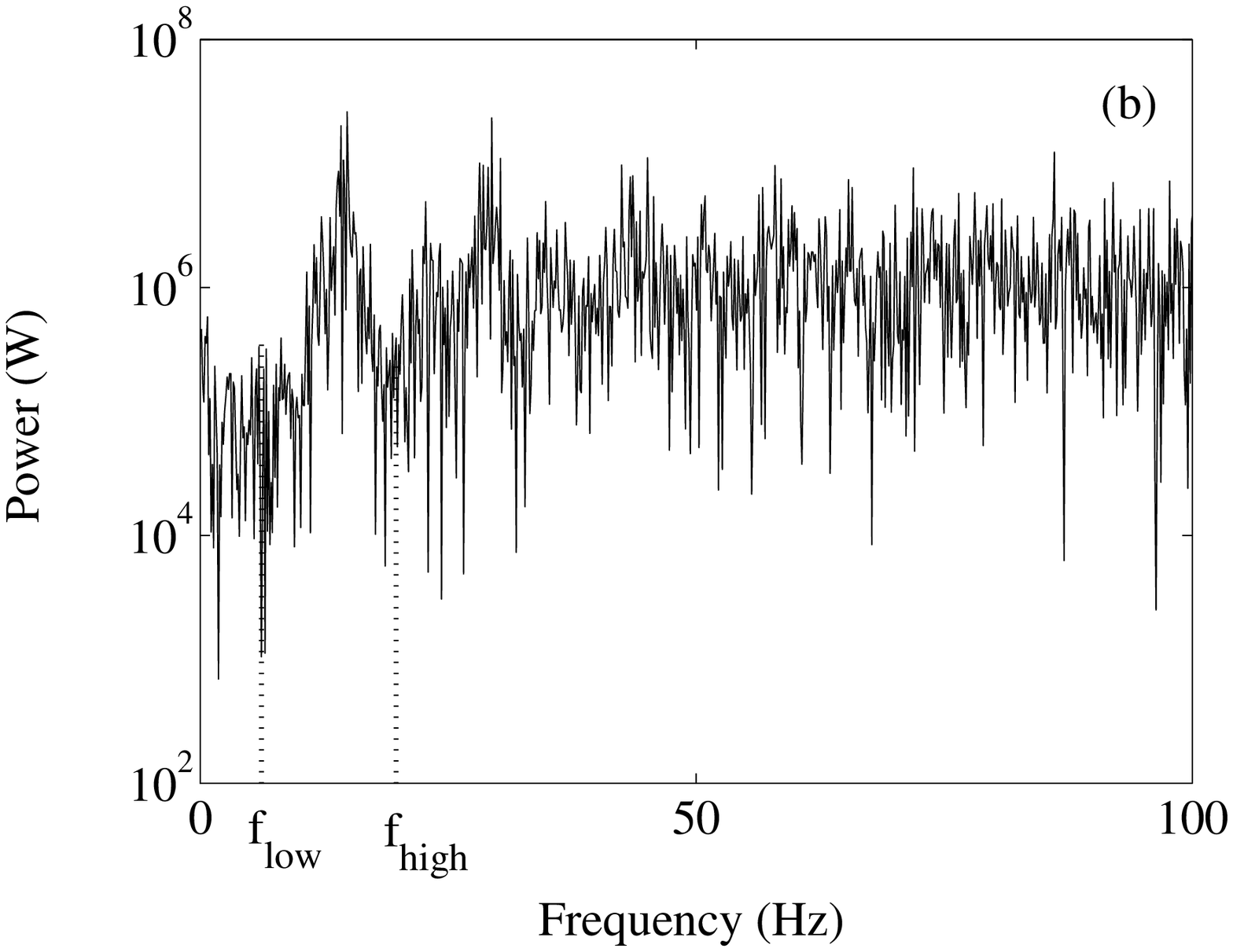}

\caption{Power spectrum of the distance from the orifice $D$ that 
perturbations form as a function of time.  Plots correspond to the 
data shown in Figure~\ref{f-Dvt}(a) in: (a) steady behavior ($+$) 
at $Q=0.336$~cm$^3$/s and (b) unsteady behavior ($\bullet$) at 
$Q=0.347$cm$^3$/s for glycerol solution. In the steady case, the 
fundamental frequency $f_0$ = 14.45~Hz, is the first 
harmonic in the power spectrum. In the unsteady case, the bandwidth
supporting the fundamental peak is larger than the steady case.}
			   
\label{f-spectra}										
\end{center}					
\end{figure}												

\begin{figure}[tbd]	
\begin{center}	

\includegraphics[width=3.1 in]{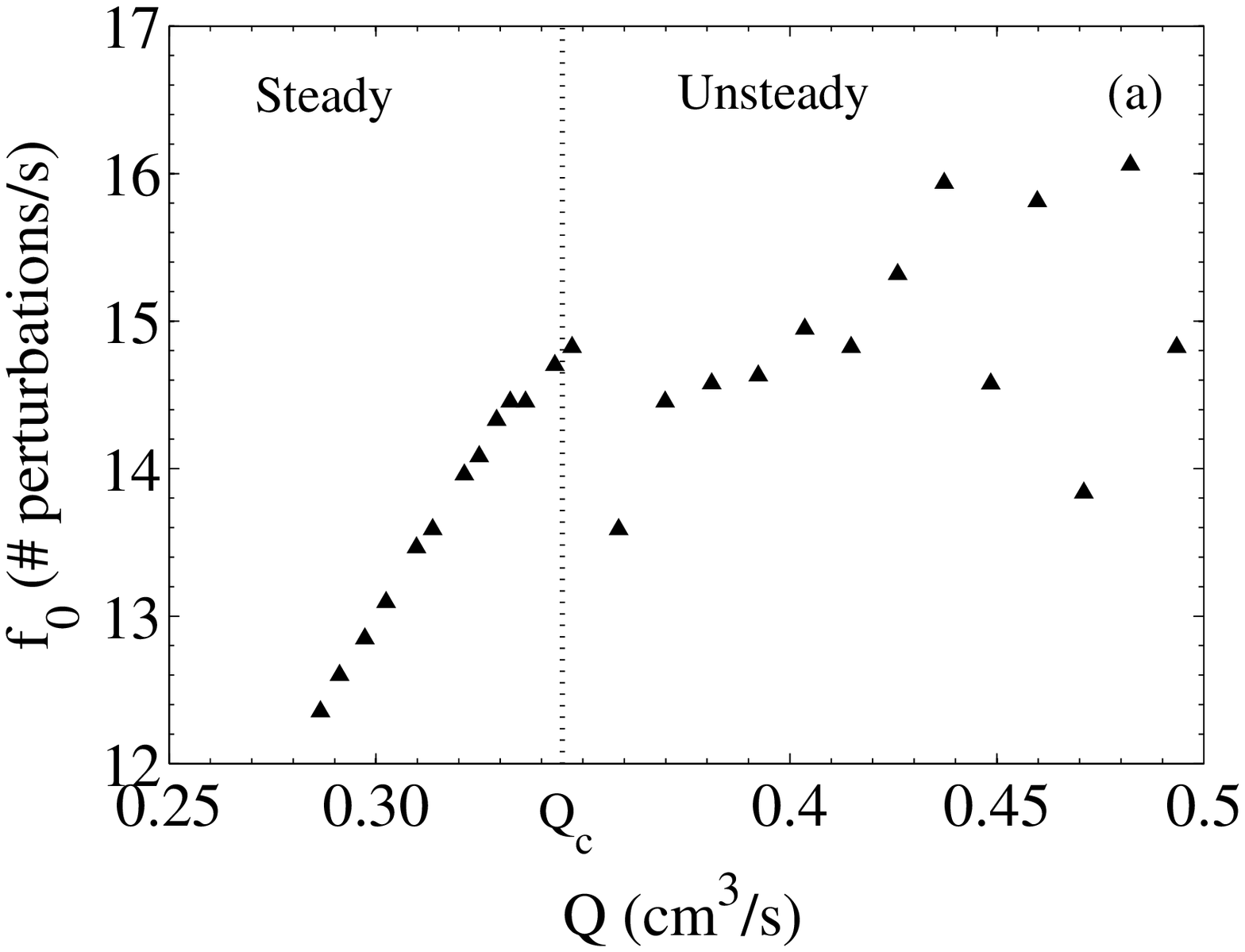}
\includegraphics[width=3 in]{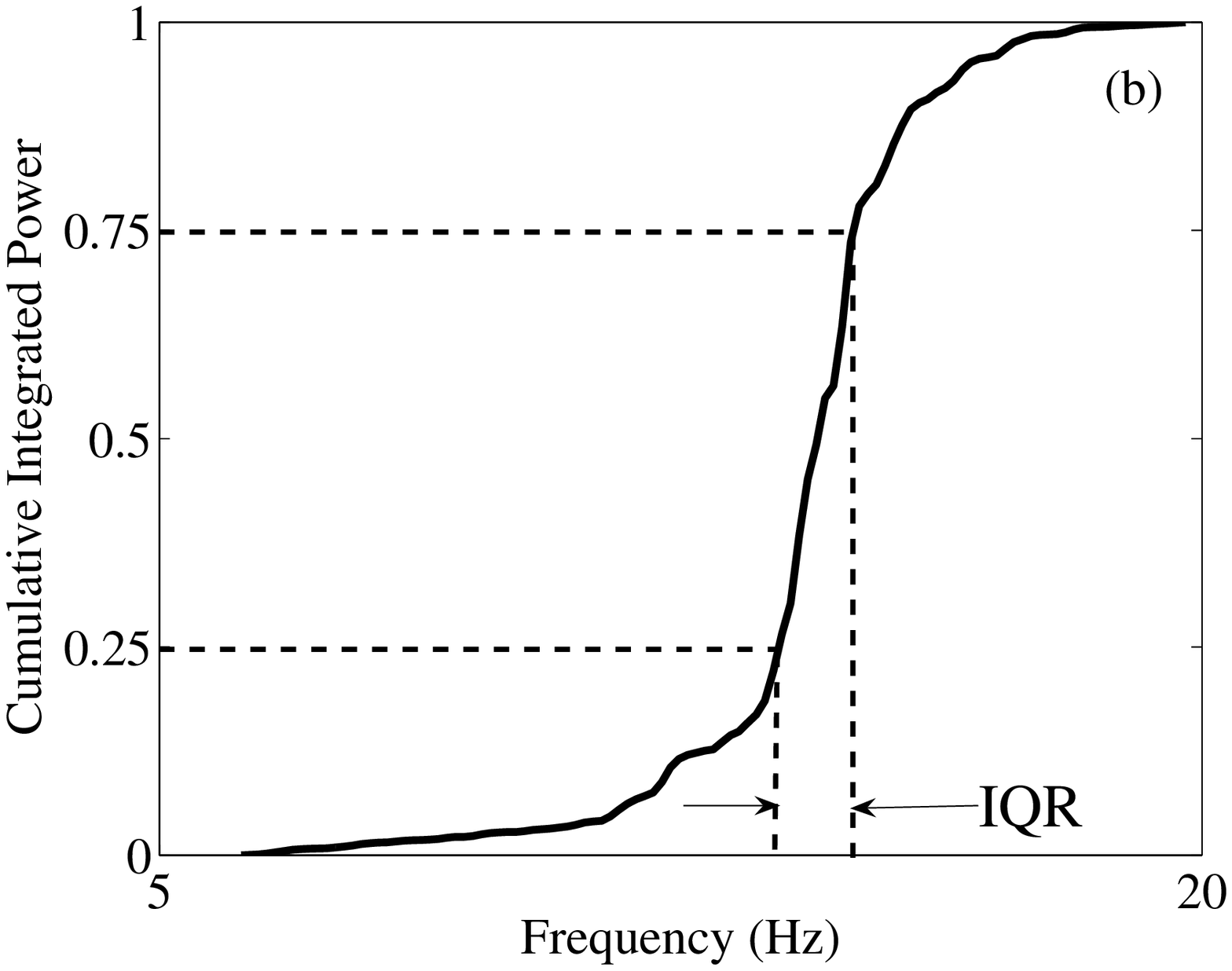}

\caption{(a) Fundamental frequency as a function of flow rate measured 
from power spectra of $D$ versus $t$ data in experiments with glycerol 
solution.  The dotted vertical line denotes the transition flow rate 
separating steady and unsteady perturbation behavior.  (b) The 
(normalized) cumulative integrated power measured within the support 
of the first peak between f$_\text{low} = 6.18$~Hz and 
f$_\text{high}=19.77$~Hz for the spectra shown in Figure~\ref{f-spectra}(b).  
The Interquartile Region (IQR) is the frequency bandwidth bounding the 
middle 50\% of the cumulative integrated power.}							
			   
\label{f-ff}															
\end{center}
\end{figure}												

\begin{figure}[tbd]	
\begin{center}	

\includegraphics[width=3 in]{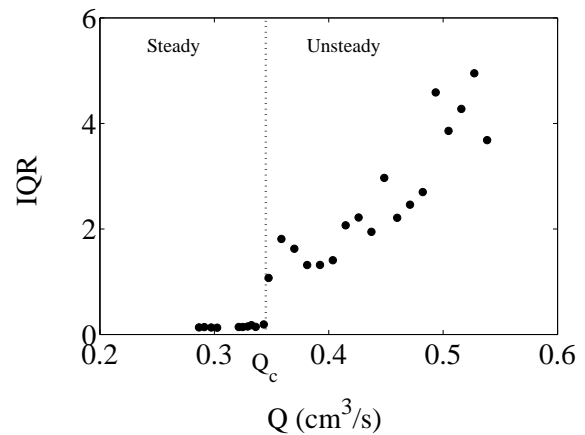}

\caption{Interquartile Region (IQR), or frequency bandwidth, as 
a function of flow rate for the glycerol solution experiments.  
An abrupt transition in the bandwidth occurs at the transition
flow rate, $Q_c$.}   
			   
\label{f-IQR}															
\end{center}
\end{figure}												

To characterize the motion of the boundary, we measure the distance 
from the orifice $D$ that each perturbation initially forms along 
the fiber (at a fixed flow rate) using edge-detection software;  a 
perturbation is detected when its amplitude ($\eta$) initially 
exceeds 1/10$^\text{th}$ of a pixel or $\approx$~0.002~cm.  The data 
shown in Figure~\ref{f-Dvt}(a) corresponds to the space-time plots 
of the unsteady ($\bullet$) and steady ($+$) experiments shown in 
Figure~\ref{f-streak}.  Figure~\ref{f-Dvt}(b) is a plot of the average 
distance from the orifice that perturbations form ($D_\text{avg}$) 
as a function of flow rate in experiments with glycerol solution.  
The vertical bars represent the standard deviation of $D$ over all
the perturbations measured at a fixed flow rate and the dotted vertical 
line represents the transition flow rate, $Q_c.$  The distance that 
perturbations form from the orifice increases monotonically with 
increasing flow rate.  In the steady case, at a given flow rate the 
distance is nearly constant, whereas in the unsteady 
case, the range of distance that perturbations form increases with 
increasing flow rate.  These results are consistent with experimental 
observations made by Duprat {\it et al.} \cite{duprat07}.

To understand the physical mechanism controlling the steady and unsteady
states we examine the power spectra of $D(t)$ in the glycerol solution 
experiments.  Figures~\ref{f-spectra}(a) and (b) represent the power 
spectra for the steady ($+$) and unsteady ($\bullet$) perturbation
behavior shown in Figure~\ref{f-Dvt}(a).  In the steady case the 
fundamental frequency ($f_0$), which is the first harmonic of the power 
spectra, represents the rate at which perturbations form along the fiber 
(e.g., $f_0 = 14.45$~perturbations/s in the experiment shown in 
Figure~\ref{f-streak}(b)); in the unsteady case the fundamental peak is 
much broader so that $f_0$ is less well defined.  As a function of 
increasing flow rate, the fundamental frequency (i.e., rate of 
perturbation formation) increases linearly when the perturbation 
dynamics are steady ($Q < Q_c$), and is scattered about 
$\approx 15$~perturbations/s when the dynamics are unsteady ($Q > Q_c$) 
(see Figure~\ref{f-ff}(a)).  Another feature in the power spectra 
distinguishing steady and unsteady behavior is in the frequency 
bandwidth supporting the fundamental peak; the bandwidth of the unsteady 
spectra is larger than the steady spectra in Figure~\ref{f-spectra}.  We 
characterize the bandwidth of the fundamental peak by measuring the 
Interquartile Region (IQR).  The IQR is defined as the frequency bandwidth 
bounding the middle 50\% of the (normalized) cumulative integrated power 
under the fundamental peak; an example is shown in Figure~\ref{f-ff}(b) 
corresponding to the power spectra in Figure~\ref{f-spectra}(b)\footnote{The 
(normalized) cumulative integrated power equals 
$\left(\int_{f_{low}}^{f(i)} P(f)\, df\right)/
\left(\int_{f_{low}}^{f_{high}} P(f)\, df\right),$ 
where  $f_{low}$ and $f_{high}$ are lower and upper frequency bounds 
supporting the fundamental peak (see Figure~\ref{f-spectra}(b)),
$f_{low} \le f(i) \le f_{high},$ and $P(f)$ is the power at frequency $f.$}. 
The IQR, or bandwidth, measures the modulation of the fundamental frequency, 
or more physically, the modulation of the rate at which perturbations form 
along the fiber.  A jump in the bandwidth occurs at the transition flow rate 
$Q_c$ in the glycerol solution experiments, as shown in Figure~\ref{f-IQR}.  
For $Q < Q_c$, the bandwidth is nearly zero, thus the rate of perturbation 
formation is nearly constant resulting in longer-time steady perturbation 
behavior. For $Q > Q_c,$ the bandwidth is sizable and increases 
with increasing flow rate, thus there is a significant modulation of the rate 
at which perturbations form.  It is this large modulation that results in the 
longer-time unsteady dynamics of the perturbations.   While the transition in 
Figure~\ref{f-IQR} is striking, it is not entirely clear whether it
is a subcritical or supercritical transition, and if subcritical, whether
the transition is hysteretic.

\section{Conclusions}
\label{sec-conc}

In an experimental study, we examine the motion of an annular
viscous film flowing under the influence of gravity down the outside 
of a vertical fiber.  We find the unperturbed flow is well 
approximated by a steady, unidirectional parallel flow when 
$\text{Re} \le 10$.  The dynamics of the perturbed flow can be 
divided into three stages: (i) initial exponential growth of the 
perturbation amplitude accompanied by a decrease in the perturbation 
wavelength; (ii) nonlinear saturation of the perturbation 
amplitude and wavelength; and (iii) longer-time behavior in which the 
perturbation wavelength may (unsteady) or may not (steady) vary along 
the film.  During the first stage, we find linear stability theory 
results developed from a long-wave Stokes flow model \cite{cras06} 
are in excellent agreement with the initial growth of perturbations 
measured in experiments.  The agreement between linear stability 
results developed from a moderate Reynolds number model 
\cite{sisoev06} and experimental data are not as strong as in 
the Stokes flow case.

A close examination of the longer-time steady and unsteady behavior 
of interfacial perturbations is shown to be correlated to the range 
of: (i) the rate of exponential growth of the perturbation amplitude; 
and (ii) the location along the fiber where perturbations initially form. 
In particular, we find the rate of growth of the amplitude and the 
location along the fiber where perturbations form is nearly constant 
for the steady case, and varies over a range of values in the unsteady 
case.  Furthermore, we find the transition in the longer-time perturbation 
dynamics from unsteady to steady behavior at a critical flow rate occurs 
because of a transition in the rate at which perturbations naturally form along 
the free surface of the film.  In the steady case, the rate of perturbation 
formation is nearly constant, resulting in the perturbations remaining 
equally spaced as they travel with the same terminal speed down the fiber.  
In the unsteady case, the rate of perturbation formation is modulated which 
results in the modulation of the initial speed and spacing between 
perturbations and ultimately leads to the coalescence of perturbations 
further down the fiber.  It is not clear whether this transition is 
subcritical or supercritical, and if subcritical, whether the transition 
is hysteretic.

\begin{acknowledgments}
We would like to thank A. Belmonte, M. G. Forest, M. Frey, D. Henderson, 
H. Segur, H. Stone and T. Witelski for many helpful discussions and 
Timothy Baker for his aid in building the experimental apparatus.  
This research was supported by a National Science Foundation REU 
grant (PHY-0097424 \& PHY-0552790).
\end{acknowledgments}

\appendix*
\section{}

Formula for the constant coefficients described in the dispersion 
relation derived by Sisoev {\it et al.} \cite{sisoev06} are:
\begin{widetext}
\begin{subequations}
\begin{eqnarray}
a_{0,0} &=& -{1 \over 5 \delta_{\eps}}
\left(1 + {\eps \over 2}\right)(3 + \phi_1),
\qquad a_{0,3} = {1 \over 5 \delta_{\eps}}\left(1 + {\eps \over 2}\right),\\
a_{0,1} &=& {b_0 \over 16\eps^5(\phi(\eps))^2}
\left({b_1 \over b_0} - 2\phi_1-6\right)-{1 \over 5 \delta_{\eps}}
\left(1 + {\eps \over 2}\right)\left(\eps \over \kappa(1+\eps)\right)^2,\\
a_{1,1} &=& {2 b_0 \over 16\eps^5(\phi(\eps))^2}, \qquad
a_{1,0} = {1 \over 5 \delta_{\eps}}\left(1 + {\eps \over 2}\right),
\end{eqnarray} 
where
\begin{eqnarray}
b_0 &=& 2(1+\eps)^6(\ln(1+\eps))^2+(2-3(1+\eps)^2)(1+\eps)^4\ln(1+\eps)
\nonumber\\
&+&{17 \over 12}(1+\eps)^6-{5 \over 2}(1+\eps)^4+{5 \over 4}(1+\eps)^2 
-{1 \over 6},\\
b_1 &=& {17 \eps \over 2}(1+\eps)^5-10\eps(1+\eps)^3
+{5 \eps \over 2}(1+\eps)+4\eps(1+\eps)^5(\ln(1+\eps))^2
\left(3+ {1 \over \ln(1+\eps)}\right) \nonumber\\
&+& \eps(2-3(1+\eps)^2)(1+\eps)^4\ln(1+\eps)\left({4 \over 1 + \eps}
-{6(1+\eps) \over 2 -3(1+\eps)^2} + {1 \over (1+\eps)\ln(1+\eps)}\right),\\
\phi_1 &=& {\eps(1+\eps)^3(4\ln(1+\eps)+1)-\eps -7\eps^2-9\eps^3-3\eps^4 \over
(1+\eps)^4\ln(1+\eps)-\eps-{7 \over 2}\eps^2-3\eps^3-{3 \over 4}\eps^4}-3,\\
\phi(\eps) &=& {4(1+\eps)^4\ln(1+\eps)-((3(1+\eps)^2-1)((1+\eps)^2-1))
\over 16 \eps^3},\\
\delta_{\eps} &=& 9 \delta(\phi(\eps))^2, \qquad
\delta = {1 \over 45 \nu^2}\left(\rho g^4 h_0^{11} \over \sigma\right)^{1/3},
\qquad \kappa = \left(\rho g h_0^2 \over \sigma\right)^{1/3}.
\end{eqnarray}
\end{subequations}
\end{widetext}


\begin{thebibliography}{99}

\bibitem{quere99} D. Quere, Annu.~Rev.~Fluid~Mech. {\bf 31}, 347 (1999).

\bibitem{ruck02} E. Ruckenstein, J.~Coll.~Int.~Sci. {\bf 246}, 393 
(2002).

\bibitem{deryck96} A. de Ryck, D. Quere, J.~Fluid~Mech. {\bf 311}, 
219 (1996).

\bibitem{deryck98} A. de Ryck, D. Quere, Langmuir {\bf 14}, 1911 
(1998).

\bibitem{shen02}  A. Q. Shen, B. Gleason, G. H. McKinley, H. A. Stone,
Phys.~Fluids {\bf 14}, 4055 (2002).

\bibitem{goren62} S. L. Goren, J.~Fluid~Mech. {\bf 12}, 309 (1962).

\bibitem{quere90} D. Quere, Europhys.~Lett. {\bf 13}, 721 (1990).

\bibitem{chang99} H-C Chang, E. A. Demekhin, J.~Fluid~Mech. {\bf 380}, 
233 (1999).

\bibitem{frenk92} A. L. Frenkel, Europhys.~Lett. {\bf 18}, 583 (1992).

\bibitem{kall94} S. Kalliadasis, H.-C. Chang, J.~Fluid~Mech. {\bf 261}, 
135 (1994).

\bibitem{lin75} S. P. Lin, W. C. Liu, AIChE~J. {\bf 21}, 775 (1975).

\bibitem{lister06} J. R. Lister, J. M. Rallison, A. A. King,
L. J. Cummings, O. E. Jensen, J.~Fluid~Mech. {\bf 552}, 311 (2006).

\bibitem{zucch05}  S. Zuccher, Exp.~Fluids {\bf 39}, 694 (2005).

\bibitem{kliak01} I. L. Kliakhandler, S. H. Davis, S. G. Bankoff, 
J.~Fluid~Mech. {\bf 429}, 381 (2001).

\bibitem{cras06} R. V. Craster, O. K. Matar, J.~Fluid~Mech. {\bf 553}, 
85 (2006).

\bibitem{solo87} F. J. Solorio, M. Sen, J.~Fluid~Mech. {\bf 183}, 
365 (1987).

\bibitem{trif92} Y. Y. Trifonov, AIChE~J. {\bf 38}, 821 (1992).

\bibitem{sisoev06}  G. M. Sisoev, R. V. Craster, O. K. Matar, S. V.
Gerasimov, Chem.~Eng.~Sci. {\bf 61}, 7279 (2006).

\bibitem{duprat07} C. Duprat, C. Ruyer-Quil, S. Kalliadasis, 
F. Giorgiutti-Dauphine, Phys.~Rev. Lett. {\bf 98}, 244502 (2007).

\bibitem{plat73} J. Plateau, {\it Statique experimentale et theorique
des liquides soumis aux seules forces molecularies}, 
(Gauthier-Villars, Paris, 1873).

\bibitem{ray1879} W.S. Rayleigh, Proc.~Lond.~Math.~Soc. {\bf 10} 4 (1879).

\bibitem{ray1892} W.S. Rayleigh, Philos.~Mag. {\bf 34} 145 (1892).

\bibitem{weber31} C. Weber, Z. angew. Math. Mech. {\bf 11}, 136 
(1931).

\bibitem{chan61} S. Chandrasekhar, {\it Hydrodynamic and Hydromagnetic
Stability}, (Dover, New York, 1961).

\bibitem{donn66} R. J. Donnelly, W. Glaberson, Proc.~Roy.~Soc.~A {\bf 290}, 
547 (1966).

\bibitem{smolka06} L. Smolka, J. North, B. Guerra, Bull.~Am.~Phys.~Soc. 
{\bf 51}, GD.00006 (2006).

\bibitem{scor90}  J. Soria, W. K. Chiu, M. P. Norton, 
Exp.~Therm.~Fluid~Sci. {\bf 3} 291 (1990).

\bibitem{frei77} W. Frei \& C. C. Chen, IEEE Transactions on Computers 
{\bf 26}, 988 (1977).

\bibitem{shka04}  V. Ya. Shkadov, Fluid~Dyn. {\bf 2}, 29 (1967).

\bibitem{keller73} J. B. Keller, S. I. Rubinow, Y. O. Tu, Phys.~Fluids 
{\bf 16}, 2052 (1973).

\end{thebibliography}
\end{document}